\documentclass[12pt,preprint]{aastex}

\newcommand{\OM}      {${\Omega}_{\mathrm M}$}
\newcommand{\OL}      {${\Omega}_{\Lambda}$}
\newcommand{\OX}      {${\Omega}_{\mathrm X}$}
\newcommand{\OT}      {${\Omega}_{\mathrm total}$}

\newcommand{\snia}    {SN~Ia}
\newcommand{\sneia}   {SNe~Ia}

\usepackage{graphicx}
\usepackage{longtable}

\makeatletter

\usepackage{natbib}

\makeatother
\begin{document}

\title{The ESSENCE Supernova Survey: Survey Optimization, Observations, and Supernova Photometry}

\author{
{G.~Miknaitis}\altaffilmark{1}, 
{G.~Pignata}\altaffilmark{2}, 
{A.~Rest}\altaffilmark{3}, 
{W.~M.~Wood-Vasey}\altaffilmark{4}, 
{S.~Blondin}\altaffilmark{4}, 
{P.~Challis}\altaffilmark{4}, 
{R.~C.~Smith}\altaffilmark{3}, 
{C.~W.~Stubbs}\altaffilmark{4,5},
{N.~B.~Suntzeff}\altaffilmark{3,6}, 
{R.~J.~Foley}\altaffilmark{7}, 
{T.~Matheson}\altaffilmark{8}, 
{J.~L.~Tonry}\altaffilmark{9}, 
{C.~Aguilera}\altaffilmark{3}, 
{J.~W.~Blackman}\altaffilmark{10}, 
{A.~C.~Becker}\altaffilmark{11}, 
{A.~Clocchiatti}\altaffilmark{2}, 
{R.~Covarrubias}\altaffilmark{11},
{T.~M.~Davis}\altaffilmark{12}, 
{A.~V.~Filippenko}\altaffilmark{7}, 
{A.~Garg}\altaffilmark{4,5}, 
{P.~M.~Garnavich}\altaffilmark{13}, 
{M.~Hicken}\altaffilmark{4,5}, 
{S.~Jha}\altaffilmark{7,13}, 
{K.~Krisciunas}\altaffilmark{13,6},
{R.~P.~Kirshner}\altaffilmark{4},
{B.~Leibundgut}\altaffilmark{14}, 
{W.~Li}\altaffilmark{7}, 
{A.~Miceli}\altaffilmark{11}, 
{G.~Narayan}\altaffilmark{4,5},
{J.~L.~Prieto}\altaffilmark{15}, 
{A.~G.~Riess}\altaffilmark{16,17},
{M.~E.~Salvo}\altaffilmark{10}, 
{B.~P.~Schmidt}\altaffilmark{10}, 
{J.~Sollerman}\altaffilmark{12, 18}, 
{J.~Spyromilio}\altaffilmark{14}, 
{A.~Zenteno}\altaffilmark{3}
}
\email{gm@fnal.gov}

\altaffiltext{1}{Fermi National Accelerator Laboratory, P.O. Box 500, Batavia, IL 60510-0500}
\altaffiltext{2}{Pontificia Universidad Cat\'olica de Chile, Departamento de Astronom\'ia y Astrof\'isica, Casilla 306, Santiago 22, Chile}
\altaffiltext{3}{Cerro Tololo Inter-American Observatory, National Optical Astronomy Observatory, Casilla 603, La Serena, Chile}
\altaffiltext{4}{Harvard-Smithsonian Center for Astrophysics, 60 Garden Street, Cambridge, MA 02138}
\altaffiltext{5}{Department of Physics, Harvard University, 17 Oxford Street, Cambridge, MA 02138}
\altaffiltext{7}{Department of Astronomy, 601 Campbell Hall, University of California, Berkeley, CA 94720-3411}
\altaffiltext{8}{National Optical Astronomy Observatory, 950 North Cherry Avenue, Tucson, AZ 85719-4933}
\altaffiltext{9}{Institute for Astronomy, University of Hawaii, 2680 Woodlawn Drive, Honolulu, HI 96822}
\altaffiltext{10}{The Research School of Astronomy and Astrophysics, The Australian National University, Mount Stromlo and Siding Spring Observatories, via Cotter Road, Weston Creek, PO 2611, Australia}
\altaffiltext{11}{Department of Astronomy, University of Washington, Box 351580, Seattle, WA 98195-1580}
\altaffiltext{12}{Dark Cosmology Centre, Niels Bohr Institute, University of Copenhagen, Juliane Maries Vej 30, DK-2100 Copenhagen \O, Denmark}
\altaffiltext{13}{Department of Physics, University of Notre Dame, 225 Nieuwland Science Hall, Notre Dame, IN 46556-5670}
\altaffiltext{13}{Kavli Institute for Particle Astrophysics and Cosmology, Stanford Linear Accelerator Center, 2575 Sand Hill Road, MS 29, Menlo Park, CA 94720}
\altaffiltext{6}{Department of Physics, Texas A\&M University, College Station, TX 77843-4242}
\altaffiltext{14}{European Southern Observatory, Karl-Schwarzschild-Strasse 2, D-85748 Garching, Germany}
\altaffiltext{15}{Department of Astronomy, Ohio State University, 4055 McPherson Laboratory, 140 West 18th Avenue, Columbus, OH 43210}
\altaffiltext{16}{Space Telescope Science Institute, 3700 San Martin Drive, Baltimore, MD 21218}
\altaffiltext{17}{Johns Hopkins University, 3400 North Charles Street, Baltimore, MD 21218}
\altaffiltext{18}{Department of Astronomy, Stockholm University, AlbaNova, 10691 Stockholm, Sweden}

\slugcomment{Submitted to ApJ on December 7, 2006}

\begin{abstract}
  We describe the implementation and optimization of the ESSENCE
  supernova survey, which we have undertaken to measure the equation
  of state parameter of the dark energy.  We present a method for
  optimizing the survey exposure times and cadence to maximize our
  sensitivity to the dark energy equation of state parameter $w=P/\rho
  c^2$ for a given fixed amount of telescope time.  For our survey on
  the CTIO 4m telescope, measuring the luminosity distances and
  redshifts for supernovae at modest redshifts ($z\sim0.5\pm0.2$) is
  optimal for determining $w$.  We describe the data analysis pipeline
  based on using reliable and robust image subtraction to find
  supernovae automatically and in near real-time.  Since making
  cosmological inferences with supernovae relies crucially on accurate
  measurement of their brightnesses, we describe our efforts to
  establish a thorough calibration of the CTIO 4m natural photometric
  system.

  In its first four years, ESSENCE has discovered and
  spectroscopically confirmed 102 type Ia SNe, at redshifts from 0.10
  to 0.78, identified through an impartial, effective methodology for
  spectroscopic classification and redshift determination.  We present
  the resulting light curves for the all type Ia supernovae found by
  ESSENCE and used in our measurement of $w$, presented in
  \citet{wood-vasey07}.
\end{abstract}

\keywords{cosmology: observations --- supernovae --- surveys --- methods: data analysis}
\pagebreak
\section{Introduction}
\label{sec:introduction}

This is a report on the first four years of the ESSENCE survey
(Equation of State: SupErNovae trace Cosmic Expansion), a program to
measure the cosmic equation of state parameter to a precision of
$10\%$ through the discovery and monitoring of high redshift
supernovae.  The motivations and goals of ESSENCE, as well as the
methods and data are presented here.  ESSENCE is part of the
exploration of the new and surprising picture of an accelerating
universe, which has become the prevailing cosmological paradigm.  This
paradigm is supported by essentially all current observations,
including those based on supernova distances, the large scale
clustering of matter, and fluctuations in the cosmic microwave
background. The free parameters of this concordance model can
consistently fit these diverse and increasingly precise measurements.

This paper describes the survey design and optimization, and the
acquisition and photometric analysis of our data through to the
generation of photometrically calibrated SN light curves. The
companion paper by \citet{wood-vasey07} describes how luminosity
distances are measured from the SN light curves and derives
constraints on $w$ from the ESSENCE observations.

\subsection{Cosmology and Dark Energy}

While the current observational agreement on a concordance model is
surprisingly good \citep{tegmark04,eisenstein05,spergel06}, it comes
at the high cost of introducing two unknown forms of mass-energy:
non-baryonic dark matter and dark energy that exerts negative
pressure.  Each is a radical idea, and it is only because multiple
independent observations require their existence that we have come to
seriously consider new physics to account for these astronomical
phenomena.

The dark energy problem is currently one of the most challenging
issues in the physical sciences. The stark difference between the
staggeringly \emph{large} value for the vacuum energy predicted by
quantum field theory and the cosmic vacuum energy density inferred
from observations leads us to wonder how this vacuum energy of the
Universe could be so \emph{small} \citep{weinberg89, carroll92,
  padmanabhan03, peebles-ratra03}.  On the other hand the convergence
of observations that give rise to the $\Lambda$CDM concordance
cosmology, with $\Omega_\Lambda \sim 0.7$ rather than identically
zero, forces us to ask why the vacuum energy is so large.

More broadly, these cosmological observations can be interpreted as
evidence for physics beyond our standard models of gravitation and
quantum field theory. It is perhaps no coincidence that this occurs at
the friction point between these two independently successful, but as
yet unmerged paradigms. Our understanding of the gravitational
implications of quantum processes appears to be incomplete at some
level.

The dark energy problem challenges us on many fronts: theoretical,
observational and experimental. Observational cosmology has an
important role to play, and the current challenge is to undertake
measurements that will lead to a better understanding of the nature of
the dark energy \citep{detf06}.  In particular we seek to measure the
equation of state parameter $w=P/\rho c^2$ of the dark energy, as this
can help us test theoretical models. One specific goal is to establish
whether the observed accelerating expansion of the Universe is due to
a classical cosmological constant or some other new physical process.

Within the framework of Friedmann-Robertson-Walker cosmology, the only
way to reconcile the observed geometric flatness and the observed
matter density is through another component of mass-energy that does
not clump with matter.  The observation of acceleration from
supernovae is the unique clue that indicates this component must have
negative pressure
\citep{riess98,perlmutter99,riess01,tonry03,knop03,barris04,riess04,clocchiatti06,astier06,riess06}.
As the evidence from supernovae has grown more conclusive, the
intellectual focus has shifted from verifying the existence of dark
energy to constraining its properties \citep{freedman03}.
Accordingly, several large-scale, multi-year supernova surveys have
embarked on studying dark energy by collecting large, homogenous data
sets.  The Supernova Legacy Survey has published cosmological
constraints using 73 SNe from its first year sample \citep{astier06}
between redshifts 0.2 and 1.0, and it continues to accumulate data.
More recently, SDSS-II Supernova Survey \citep{frieman04} has observed
$\sim200$ supernovae at redshifts out to 0.4.  The final supernova
samples from each of these programs and ESSENCE will each number in
the hundreds.

Of the various models for dark energy currently being discussed in the
literature, the cosmological constant (i.e. some uniform vacuum energy
density) holds a special place, as both the oldest, originating with
Einstein, and, in many ways, the simplest \citep{carroll92}.  Quantum
field theory suggests how to calculate the energy of the vacuum, but
there is no plausible theoretical argument that accounts for the
small, but non-zero, value required by observations. A host of other
alternatives has been proposed, many of which appeal to slowly rolling
scalar fields, similar to those used to describe inflation. Such
models readily produce predictions that agree with the current
observational results, but suffer from a lack of clear physical
motivation, being concocted after the fact to solve a particular
problem. Another class of ideas appeals to higher dimensional ``brane
world'' physics inspired by string theory; for example, the cyclic
universe \citep{steinhardt02,steinhardt05} or modifications to gravity
due to the existence of extra dimensions \citep{dvali03}.

One straightforward way to parameterize the dark energy is by assuming
its equation of state takes the form $P=w\rho c^{2}$, where $P$ and
$\rho$ are pressure and density respectively, related by an
``equation-of-state'' parameter $w$. Non-relativistic matter has
$w=0$, while radiation has $w=+1/3$, and different proposed
explanations for the dark energy have a variety of values of $w$. In
general, to produce an accelerated expansion, a candidate dark energy
model must have $w<-1/2$ for a current matter density of \OM$\sim
1/3$.  The classical cosmological constant, $\Lambda$, of General
Relativity has an equation-of-state parameter $w=-1$ exactly, at all
times. Other models can take on a variety of effective $w$ values that
may vary with time. For example, quintessence \citep{steinhardt03}
posits a minimally coupled rolling scalar field, with an equation of
state,
\begin{equation}
w\sim\frac{P}{\rho}=\frac{\frac{1}{2}\dot{\phi}^{2}-V(\phi)}{\frac{1}{2}\dot{\phi}^{2}+V(\phi)}.
\end{equation}
In this case, the effective value of $w$ depends on the form of the
potential chosen and can evolve over time.  In general the
parameterization of dark energy in terms of $w$ is a convenient and
useful tool to compare a variety of models~\citep{weller02}.

As a first step towards determining the nature of dark energy, the
obvious place to start is to test whether the observed w is consistent
with $-1$ \citep{garnavich98}.  If not, then a cosmological constant is
ruled out as the explanation for dark energy.

If $w$ is measured to be consistent with $-1$, then while models that
exhibit an effective $w\sim-1$ are still allowed, the range in
parameter space in which they can exist will be significantly
restricted. Breaking the degeneracy between $\Lambda$ and such
``impostors'' would then require measurements of the additional
parameters that describe their time dependence.  However, the form of
such a parameterization is at present largely unrestricted and the
choice of arbitrary parameterizations influences the conclusions
derived from the analysis of the data \citep{upadhye05}.  In the
future, measurements of growth of structure, such as through weak
lensing surveys, will provide a powerful complement to supernova
measurements in constraining the properties of dark energy, as well as
checking for possible modifications to General Relativity
\citep{detf06}. In the near term, constraining $w$ under the assumption
that it is constant allows us to test a well-posed hypothesis that can
be addressed with existing facilities and methods.  While under
standard General Relativity, $w$ is bounded by the null dominant
energy condition to be greater than or equal to $-1$, we should keep
an open mind as to whether the data allow $w<-1$, since dark energy
may well arise from physics beyond today's standard theories.

Motivated by these considerations, we have undertaken a project to use
type Ia supernovae to measure $w$ with a target fractional uncertainty
of 10\%.  Observations of type Ia supernovae provided the first direct
evidence for accelerating cosmic expansion, and they remain an
incisive tool for studying the properties of the dark energy.
 
\pagebreak
\subsection{Measuring the physics of dark energy with supernovae}
\label{measuring_de}

Type Ia supernovae (\sneia) are among the most energetic stellar
explosions in the universe.  Their high peak luminosities ($4-5 \times
10^9 L_\sun$) make \sneia\ visible across a large fraction of the
observable universe. The peak luminosity can be calibrated to
$\sim15\%$ precision in flux
\citep{phillips93,hamuy96,riess96,goldhaber-stretch,guy05,jha-mlcs2k2}.
They are thus well suited to probing the expansion history during the
epoch in which the Universe has apparently undergone a transition from
deceleration to acceleration ($0<z<1$).  The utility of \sneia\ as
``standardizable'' candles was established observationally by
\citet{phillips93}, with the identification of a correlation between
peak luminosity and width of the light curves.  The ``type Ia''
designation is an observational distinction, denoting objects whose
spectra lack hydrogen or helium features, but exhibit a characteristic
absorption feature observed at $\lambda$6150, but attributed to
Si~{\sc ii}~$\lambda$6355.  These objects are now thought to be the
thermonuclear disruption of a carbon-oxygen white dwarf at or near the
Chandrasekhar mass \citep{Hoyle/Fowler:1960}, with accretion from a
companion star.  Material gained through accretion pushes the total
mass of the C-O WD above what can be supported by degeneracy pressure
and results in nuclear burning, which eventually results in a powerful
burning wave that completely destroys the star.  A large fraction of
the progenitor burns rapidly to produce $^{56}$Ni, whose radioactive
decay then powers the observed light curve \citep{Colgate/McKee:1969}.
Bolometric light curves suggest that $\sim0.7M_\odot$ of $^{56}$Ni is
produced, which suggests that the burning is incomplete
\citep{Contardo/Leibundgut/Vacca:2000,Stritzinger/etal:2006}.  There
is disagreement on important details of whether the burning wave is
supersonic (a detonation) or purely subsonic (a deflagration)
\citep{Hillebrandt/Niemeyer:2000}. Nevertheless, models for the
explosion give broad agreement with the observed light curves and
spectra, though the specifics of progenitors and explosion physics
remain unresolved
\citep{Branch/etal:1995,Renzini:1996,Nomoto/etal:2000,Livio:2000}.

Fortunately, so far the lack of a detailed understanding of supernova
physics has not prohibited the use of these objects as probes of
cosmology, as the empirical correlations of light curve shape and
color with luminosity appear to largely ``standardize'' supernovae.
Subtle effects such as how supernovae are connected to stellar
populations and how those populations may change with time and
chemical composition will certainly become important in the future as
we attempt to place ever tighter constraints on dark energy
\citep{hamuy00,jha-thesis,gallagher05,sullivan06}.  For example,
observations suggest that the brightest \sneia\ are found only in
galaxies with current star formation.

As in classical physics, the flux density from a cosmological source
falls off as the inverse square of distance,
\begin{equation}
{\mathcal{F=}}\frac{L}{4\pi D_{l}^{2}}.
\end{equation}
However, this luminosity distance, $D_{l}$, depends upon how the
universe expands as a photon travels from emitter to receiver, which
in turn depends sensitively on the composition and properties of the
constituents of the cosmic mass-energy density. Specifically, for a
flat universe the luminosity distance, $D_{l}(z)$, is given by

\begin{equation}
  D_{l}=\frac{c(1+z)}{H_{0}}\int_{0}^{z}\frac{1}
  {\sqrt{(1-\Omega_M)(1+z')^{3(1+w)}+\Omega_M(1+z')^{3}}}dz',
\end{equation}

where $w$ is taken here to be constant. In cosmological analyses, the
combination of the Hubble constant and the intrinsic luminosity of
\sneia\ is a multiplicative nuisance parameter which scales distance
measurements at all redshifts by the same amount. Thus, under the
assumption of flatness, \OM$+$\OX$=1$, when measuring $w$, the only
other free cosmological parameter is the matter density, \OM.

\begin{figure}[th]
  \plotone{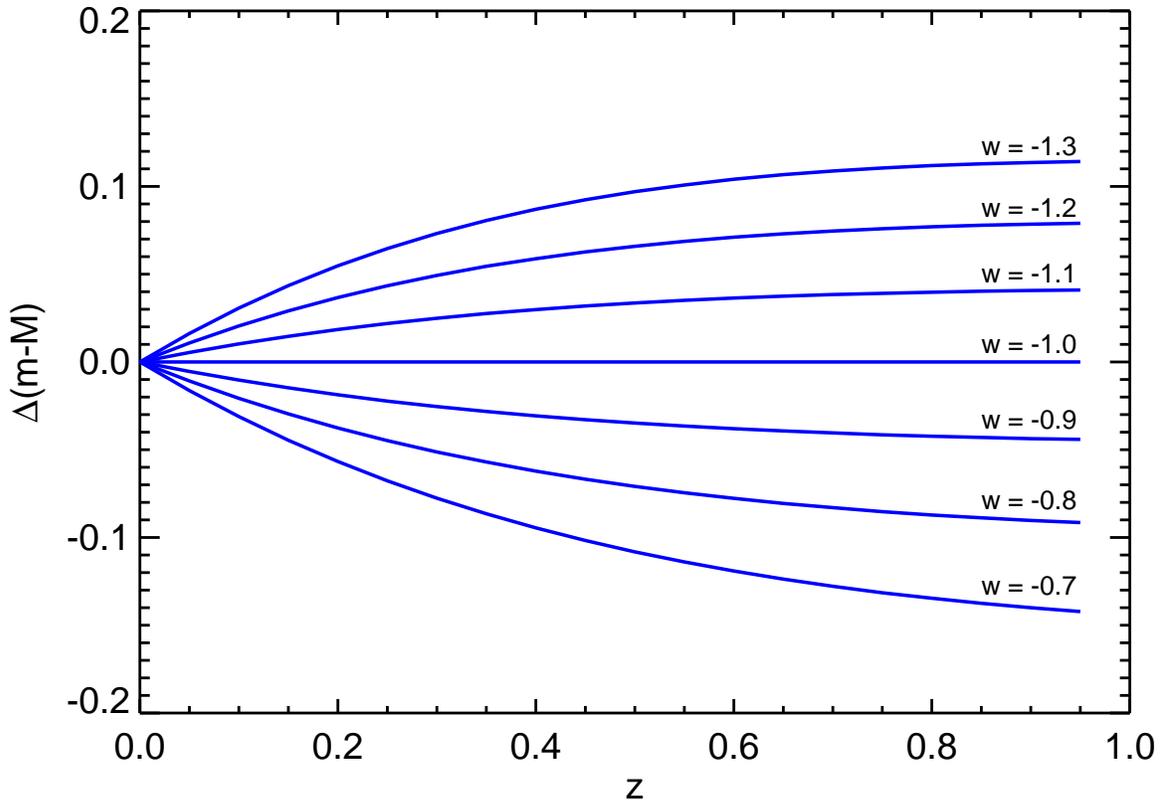}
  \caption{\label{fig:deltamu} Differences in distance modulus for
    different values of $w$ as a function of redshift, relative to
    $w=-1$, for $\Omega_{m}=0.3,\Omega_{\Lambda}=0.7$. Note that even at
    modest redshifts there is a significant fraction of the total
    asymptotic signal available.}
\end{figure}

If we seek to constrain $w$ using the luminosity distance-redshift
test, it is worth considering which redshifts are most incisive. The
relative differences in distance modulus as a function of redshift,
for different values of $w$, are shown in Figure \ref{fig:deltamu},
where \OM and \OL have been fixed at 0.3
and 0.7 respectively. 

There is a significant $w$-dependent signal even at intermediate
redshifts ($z\sim0.4$), at which observations with a 4-meter class
telescope can readily yield many supernovae each month. Of course,
observations such as the ESSENCE survey actually produce a complex set
of constraints in cosmological parameter space, but much of the signal
of interest is readily accessible at intermediate redshifts, between
0.3 and 0.8.

\subsection{Considerations for optimally constraining $w$ with \sneia\
  observations}

We wish to determine the optimal use of the time allocated for the
ESSENCE survey on the Blanco 4m for constraining $w$.  For a
ground-based survey, a variety of factors determine the number of
useful supernovae monitored, and the uncertainties associated with
each data point on the light curve. The overall quality of each
supernova light curve, in turn, determines the precision of its
luminosity distance.  Some of factors which impact the ability of a
particular survey strategy to constrain cosmological parameters
include:

\begin{itemize}
\item \emph{Typical site conditions:} Seeing, weather, sky background,
  atmospheric transmission.
\item \emph{System throughput vs. wavelength:} Aperture, optics, field
  of view, detector quantum efficiency.
\item \emph{Temporal constraints}: Telescope scheduling constraints,
  camera readout time.
\item \emph{SNR considerations}: Requisite S/N ratio and cadence
  required for distance determination.
\item \emph{Passband considerations}: Number of bands needed for
  extinction and SN color discrimination.
\item \emph{Spectroscopic considerations:} Location, availability and
  scheduling of followup spectroscopic resources
\end{itemize}

In order to optimize the observational survey strategy for ESSENCE, we
tried to parameterize several of the factors above and balance them to
obtain the strongest constraints on $w$. With the strong cosmological
signal available in the redshift range from 0.3 to 0.8, it is clear
that a wide-field camera on a 4m-class telescope can provide the
needed balance of photometric depth (\sneia\ have m$\sim$22 at peak at
z=0.5) and sky coverage.  Smaller fields of view on larger telescopes
are better suited to going to higher redshifts, while wider fields on
smaller telescopes are only able to reach redshifts where the
cosmological signal is small.  Combining these criteria with the range
of spectroscopic followup facilities available to our collaboration,
we quickly focused our analyses on the Blanco 4m telescope at CTIO
together with the MOSAIC camera as providing an optimal combination of
site (seeing plus weather), aperture, field of view, and telescope
scheduling.

Beyond the selection of appropriate telescopes and instrumentation,
there are relatively few ``free parameters'' controllable by the
observers.  These include the optical passbands used, the exposure
time in each passband for each field, the total number of fields
monitored, the cadence of the repeated observations, and ability to
obtain spectra for each supernova candidate.

Using existing knowledge of the distribution of supernova magnitudes
and colors as a function of redshift and time after explosion, we can
relate the exposure times in different passbands, for a given desired
signal-to-noise ratio (SNR).  The calibration of luminosity from light
curve shape is currently best understood in rest-frame B and V
passbands.  These passbands map to observer-frame R and I for
supernovae at $z\sim0.4$ (i.e. the uncertainties in k-corrections
\citep{nugent-kcorr} are small).  For supernovae at these redshifts,
observations taken in R and I, with the I band exposure time equal to
twice that in R, are sufficient to match the SNR in both bands and
measure distances to the \sneia.

While observations in a third bandpass would aid in determination of
color, and thus, the estimates of extinction in the host galaxies,
such observations would require significant additional observing time
and are not easily accommodated within our optimization of limited
observing time, photometric depth, sky coverage, and number of
resulting SNe.  Acquiring V band observations would provide a better
match to rest-frame B for low redshift supernovae, but supernovae in
our sample will be bright and have well-measured colors at these
redshifts.  Observations in the z-band would aid the color
determination at higher redshifts, but the low quantum efficiency of
of the MOSAIC CCD detectors, as well as the brightness of the night
sky in this band and the heavy fringing due to night-sky emission
lines make obtaining useful data in this band impractical.

Therefore, by limiting our strategy to R and I and demanding that I
band exposure times scale with R band exposure time, the survey
optimization problem then is reduced to considering a single free
parameter: the distribution of R band integration times across the
survey fields for a given fixed amount of telescope time.  What is the
balance between survey depth (which extends the redshifts probed) and
area (which increases the area covered each redshift slice)?

\begin{figure}[th]
\plotone{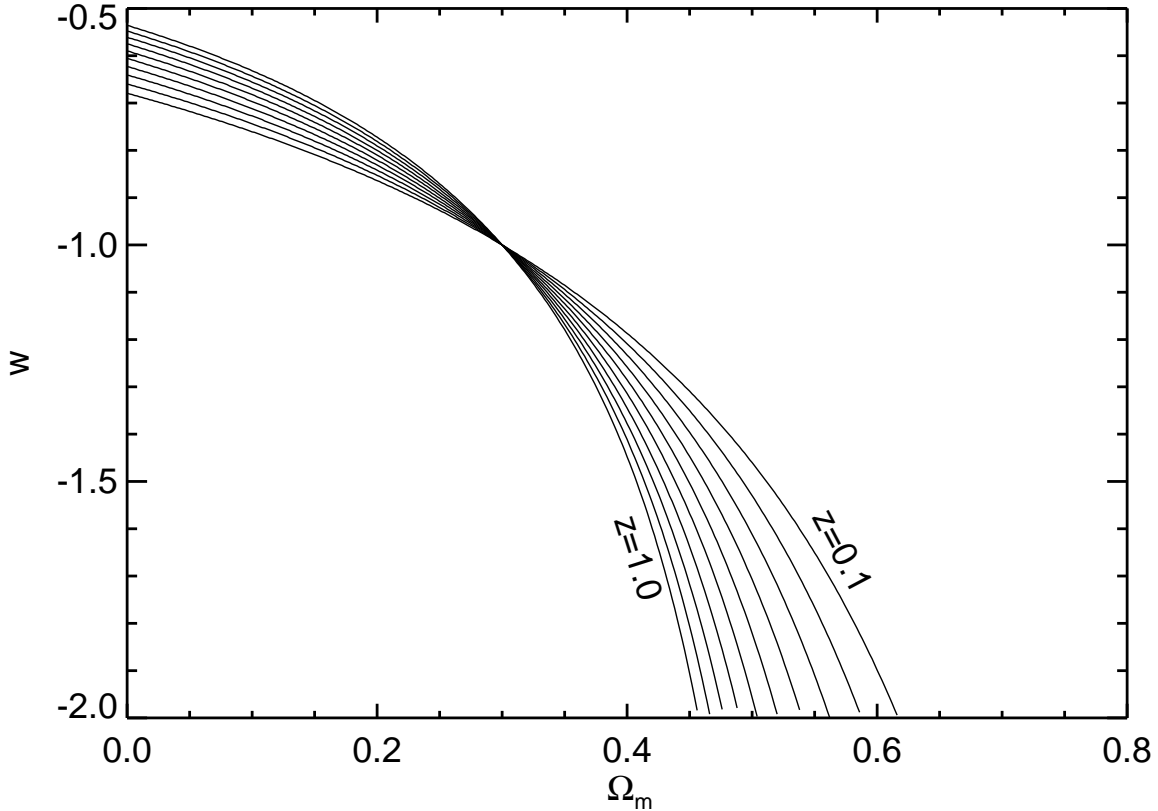}
 \caption{\label{fig:dz2omw}Curves in $\Omega_{m}$ and $w$ for perfect
   measurements of distance at redshifts from 0.1 to 1.0, in steps of
   $\Delta z=0.1$, for $\Omega_{m}=0.3,\Omega_{\Lambda}=0.7$.}
\end{figure}

Consider the cosmological information contained in a single, perfect
measurement of distance and redshift. Under the assumption of flat
geometry (and with perfect knowledge of $H_{0}$ and the intrinsic
luminosity of \sneia), each such measurement traces out a curve of
allowable values of \OM and $w$ , as shown in Figure \ref{fig:dz2omw}.
It is clear that if the goal is to measure $w$ from \sneia\ alone, a
large span in redshift is desirable in order to maximize the
orthogonality of the curves and break the degeneracy between matter
density and the equation-of-state parameter. 

However, because the difference between these curves is small even
over a large span in redshift, such a measurement would require
massive numbers of \sneia\ achievable only by next generation
experiments, such as the DES, PanSTARRS, LSST or JDEM.  In the near
term, we may appeal to other cosmological measurements to provide a
constraint on $\Omega_{m}$, such as from large scale structure
measurements.  This affords us some freedom in the redshifts at which
we make our measurements, since the constraints from distance
measurements are nearly orthogonal to an $\Omega_{m}$ prior of
$\sim$0.3 at all redshifts.  \footnote{We consider here a prior on
  $\Omega_{m}$ alone, though in reality constraints from measurements
  of the matter power spectrum, baryon acoustic oscillations or cosmic
  microwave background produce constraints which have at least mild
  degeneracy with other cosmological parameters.  This simple prior is
  sufficient for the survey optimization arguments presented here.}
Though the sensitivity to differences in cosmological models is weaker
at lower redshifts, there is a powerful observational advantage to
working there, because obtaining good photometric and spectroscopic
measurements is far cheaper in units of telescope time.

To understand the trade-offs between the cosmological sensitivity of
samples obtainable under differing observational strategies, we
carried out simulations to predict the number and distribution in
redshift and magnitude of the set of \sneia\ detectable for survey of a
given length and limiting magnitude set by the R-band exposure time.
We adopt the methodology used in \citet{tonry03} to model the
redshift-magnitude distribution of \sneia. In brief, we assume the
supernovae luminosity function used in \citet{li01a}, modeled as three
distinct luminosity classes representing ``normal'' , over-luminous
(1991T-like) and sub-luminous (1991bg-like) supernovae, each following
a Gaussian distribution. This is then convolved with an estimated
distribution of extinction due to dust in the supernova host galaxies
\citep{hatano98}.  We can then generate mock supernova samples for
various possible survey implementations.  For the purposes of survey
optimization, it is sufficient to restrict our considerations to flat
cosmologies, neglecting degeneracies with \OT.

To estimate the acheivable cosmological constraints, we use an
analytic description of how the uncertainty in distance modulus
depends on redshift, as the typical signal-to-noise ratio of the
photometry decreases at higher redshift, but the temporal sampling (in
the SN rest frame) improves due to time dilation.  The uncertainty in
distance modulus is approximated by the expression:

\begin{equation}
  \delta_{\mu}(z)=\frac{1.3}{SNR_{peak}}\times\sqrt{\frac{\Delta t_{obs}}{1+z}}\times\sqrt{\frac{N_{obs}}{N_{obs}-3}},\end{equation}

\noindent where $\Delta t_{obs}$ gives the time in days between
observations, $N_{obs}$ specifies the number of observations between
-10 and +15 days (relative to maximum) in the SN rest frame, and the
$N_{obs}-3$ term arises from three degrees of freedom in the fit of an
SN light curve -- time of maximum, luminosity at maximum and the width
of the light curve.

This contribution to the distance uncertainty due to observational
constraints is then summed in quadrature with the intrinsic dispersion
in type Ia peak luminosities, taken conservatively to be 0.2
magnitudes. With the resulting mock Hubble diagrams, we then can
predict the cosmological constraints obtainable for a given survey
depth.  

\subsection{The ESSENCE Strategy}

\begin{figure}[th]
\plotone{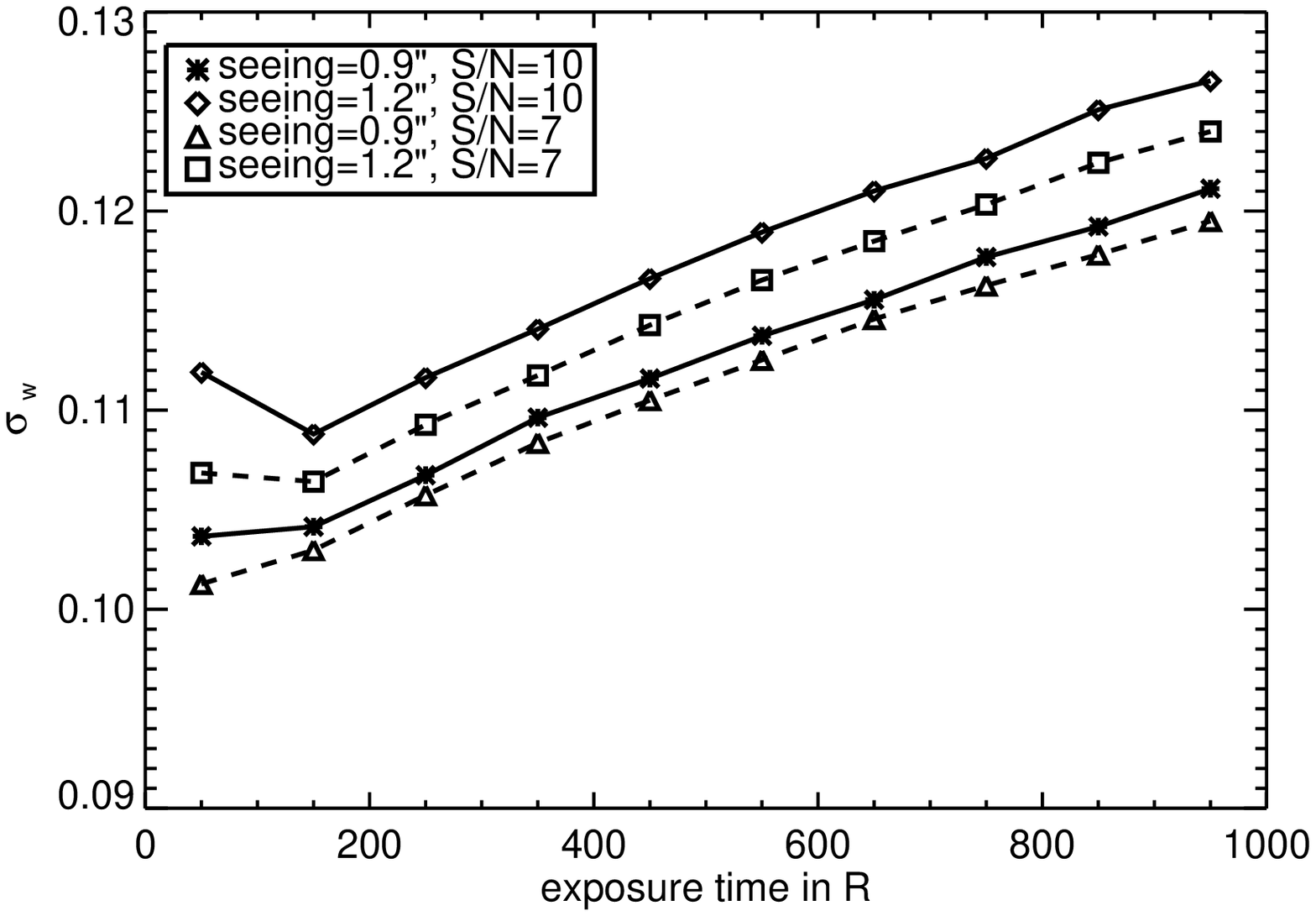}
 \caption{\label{dw_vs_exptime}Estimated final uncertainty in $w$ for
   a 5 year ESSENCE survey when combined with $\Omega_M=0.3\pm0.04$
   constraints from \citet{tegmark04}, as a function of R-band
   exposure time for the survey.  A range of typical survey seeing
   conditions and detection thresholds was chosen.  Here we show the
   effects of mean seeing, which degrades the precision of the
   photometry, and the signal-to-noise threshold at which we are able
   to detect supernovae in our data, which affects the total number
   observable.}
\end{figure}

This generalized analysis can now be applied to our selected
observational system, the Blanco 4m, in order to derive an optimal
balance of photometric depth (or equivalently exposure time) and sky
coverage given the range of conditions one might expect during a
survey using a fixed amount of observing time.  We assumed a five year
survey with approximately 15 nights per year spread over three months
each year.  The results are shown in Figure \ref{dw_vs_exptime}.  We
find that the final achievable uncertainty in $w$ is surprisingly
insensitive to the survey depth, with the trade-off between the number
of supernovae and the redshifts at which they are found roughly
cancelling.  There is a weak optimum at $t_{R}=200$ seconds because
very shallow surveys lose cosmological leverage as the redshift range
probed decreases. After initially opting for a range of exposure times
designed to match a range of redshift bins covering $z=0.3 -- 0.8$ in
2002, and finding that the efficiency at shorter exposure times was
inadequate, we settled on exposure times of $t_{R}=200$ seconds and
$t_{I}=400$ seconds as the baseline for the rest of the ESSENCE
survey.

\pagebreak

\section {The ESSENCE Survey}
\label{sec:observations}

\subsection {Observations}

Based on the survey strategy described above, the ESSENCE team
submitted a proposal to the NOAO Survey program in 2002.  We chose to
propose a survey strategy to share time with the ongoing SuperMACHO
survey, which uses only half nights on the Blanco telescope. ESSENCE
was awarded 30 half nights per year for a five year program (recently
extended to six), as well as additional calibration time on the CTIO
0.9m telescope together with some followup time on the WIYN 3.5m
telescope.  The ESSENCE time is generally scheduled during dark and
grey time for three consecutive months, from October to December each
year, although the timing of new moons sometimes moves the schedule
into September or January.  Each month, we observe every other night
over a span of 20 days centered on new moon.  This schedule leaves
approximately ten bright nights each month with no light curve
coverage.

\subsubsection{The Instrument}

ESSENCE survey data are taken using the MOSAIC II imaging camera,
which consists of eight 2048x4096 pixel charge-coupled devices (CCDs)
arranged in two rows of four, with gaps corresponding to approximately
50 pixels between rows and 35 pixels between columns.  In the f/2.87
beam at prime focus, this yields a field of view of 0.6 degrees on a
side for a total area of 0.36 square degrees on the sky.  The CCDs are
thinned back-illuminated silicon devices manufactured by SiTE with
15-um pixels.  At the center of the focal plane, each pixel subtends
0.27 arc-seconds on a side, though the pixel scale varies
quadratically as a function of radius due to optical aberrations, such
that pixels at the corners of the camera subtend a smaller area on the
sky by 8\%.

The CCDs are read out in dual-channel mode, in which the chip is
bisected in the long direction and read out in parallel through two
separate amplifiers, for a read time of about 100 seconds. Because the
amplifiers are not perfectly identical, we treat the sixteen resultant
1048x4096 ``amplifier images'' as independent data units in our data
reduction. All ESSENCE observations are taken through the Atmospheric
Dispersion Corrector (ADC), which is composed of two independently
rotating prisms that compensate for variation in atmospheric
refraction with airmass.

\subsubsection {ESSENCE Fields}
We selected fields that are equatorial, so that they can be accessed
by telescopes in the northern and southern hemisphere for followup
spectroscopy.  The fields are spaced across the sky so that all
observations may be taken at low airmass.  We chose regions with low
Milky Way extinction, for maximum visibility of these faint
extra-galactic sources and to minimize systematic error incurred by
correcting for extinction due to the Milky Way.  Fields with
contamination from bright stars, whose large footprint in the imaging
data would reduce the effective search area, were avoided.  Additional
considerations in field selection included a preference for areas with
minimal IR cirrus (based on IRAS maps), a preference for areas out of
both the galactic and ecliptic planes, and a preference for fields
which overlapped previous wide-field surveys (such as the Sloan
Digital Sky Survey (SDSS), NOAO Deep Wide-Field Survey, and the Deep
Lens Survey).

The first ESSENCE observations commenced on 2002 September 28. For
this first year of operations, a set of 36 fields was defined.  These
fields were divided into two sets, which were then observed every
other ESSENCE night, resulting in an cadence of every 4 nights on any
particular field. This proved to be a challenging inaugural season for
the project. The \emph{El Ni\~no} Pacific weather pattern was in
effect, which produced heavy cloud cover much of time, resulting in
either lost observing time or data of such poor quality that the
detection of faint supernovae was often not possible. Also, the newly
commissioned computing cluster experienced catastrophic failure
shortly after data collection began, bringing real-time analysis of
the data to a standstill for much of that observing campaign.  On the
night of November 9, the I-band filter sustained significant damage,
resulting in a crack.  This severely degraded the I band data quality
in CCDs 1 and 2 (amplifiers 1-4), resulting in a diminished effective
field of view for the rest of the season. This filter was replaced on
May, 24, 2003.  

As described below, many of the 2002 fields have not yet been repeated
to provide template images to extract the supernova light curves.  The
complete analysis of the 2002 data will take place when these
reference images are obtained.  We provide summary information about
the 15 spectroscopically confirmed Ia from this season in Table
\ref{snid_table}, we only present the light curves for four of these
objects for which current reductions are of sufficient quality to
merit use in the cosmological analysis in \citet{wood-vasey07}.  The
final ESSENCE supernova sample will include all of the 2002 objects.

Observations for the second year of ESSENCE began on September 28,
2003.  In order to facilitate scheduling of follow-up observations
with the Hubble Space Telescope (HST), which requires advance
knowledge of the approximate location of the targets, it was necessary
to cluster the search fields together into four groups.  The new field
set consisted of 32 fields, clustered spatially in sets of eight, such
that they were within the pointing error box of the HST.  To the extent
possible, fields from 2002 were used as the basis for the new fields.
The fields were again divided into two separate sets, observed on
alternating nights, providing for an observational cadence of every 4
nights for any given field.  In table \ref{fieldtable}, we list the
coordinates of the 32 search fields monitored by ESSENCE from 2003
onward.  Results from the subset of nine ESSENCE supernovae observed
with HST were presented in \citet{krisciunas05}.

\begin{deluxetable}{lrr}
  \tablewidth{0pc} \tablecaption{Coordinates of the centers of the
    ESSENCE search fields.
\label{fieldtable}
}
\tablehead{
\colhead{Field Name} &    \colhead{RA (J2000)} & \colhead{Dec (J2000)} \\
}
\startdata
waa1  & 23:29:52.92 &	-08:38:59.7 	  		 \\
waa2  & 23:27:27.02 &	-08:38:59.7 	  		 \\
waa3  & 23:25:01.12 &	-08:38:59.7 	  		 \\
waa5  & 23:27:27.02 &	-09:14:59.7 	  		 \\
waa6  & 23:25:01.12 & 	-09:14:59.7  	   	 	 \\
waa7  & 23:30:01.20 &	-09:44:55.9 	  		 \\
waa8  & 23:27:27.02 &	-09:50:59.7 	  	  	 \\
waa9  & 23:25:01.12 &	-09:50:59.7 	  		 \\
wbb1  & 01:14:24.46 &	00:51:42.9 	  		 \\
wbb3  & 01:09:36.40 &	00:46:43.3 	  		 \\
wbb4  & 01:14:24.46 &	00:15:42.9 	  		 \\
wbb5  & 01:12:00.46 &	00:15:42.9 	  		 \\
wbb6  & 01:09:00.16 &	+00:10:43.3 	  		 \\
wbb7  & 01:14:24.46 &	-00:20:17.1 	  		 \\
wbb8  & 01:12:00.46 &	-00:20:17.1 	  		 \\
wbb9  & 01:09:36.40 &	-00:25:16.7 	  		 \\
wcc1  & 02:10:00.90 &	-03:45:00.0 	  		 \\
wcc2  & 02:07:40.60 &	-03:45:00.0 	  		 \\
wcc3  & 02:05:20.30 &	-03:45:00.0 	  		 \\
wcc4  & 02:10:01.20 &	-04:20:00.0 	  		 \\
wcc5  & 02:07:40.80 &	-04:20:00.0 	  		 \\
wcc7  & 02:10:01.55 &	-04:55:00.0 	  		 \\
wcc8  & 02:07:41.03 &	-04:55:00.0 	  		 \\
wcc9  & 02:05:20.52 &	-04:55:00.0 	  		 \\
wdd2  & 02:31:00.25 &	-07:48:17.3 	  		 \\
wdd3  & 02:28:36.25 &	-07:48:17.3 	  		 \\
wdd4  & 02:34:30.35 &	-08:19:18.2 	  		 \\
wdd5  & 02:31:00.25 &	-08:24:17.3 	  		 \\
wdd6  & 02:28:36.25 &	-08:24:17.3 	  		 \\
wdd7  & 02:33:24.25 &	-08:55:18.2 	  		 \\
wdd8  & 02:31:00.25 &	-09:00:17.3 	  		 \\
wdd9  & 02:28:36.25 &	-09:00:17.3 	  		 \\
\enddata
\tablecomments{For reference, the CTIO 4-m MOSAIC II detector has a
  field of view of 0.36 square degrees.  }
\end{deluxetable}

\clearpage

Weather and observing conditions for 2003 were greatly improved over
2002, though still somewhat sub-standard for typical conditions at
Cerro Tololo.  Unfortunately, one of the MOSAIC CCDs (containing
amplifiers 5 and 6) failed shortly before the observations began,
resulting in a 12.5\% loss in efficiency.  The failed CCD was replaced
before our 2004 observing season, allowing us to recover the lost
efficiency from then on.  For the third year and fourth years of
ESSENCE, we maintained the same set of fields as in 2003 and the
MOSAIC imager was stable.  The supernovae yields for each of the four
years of the survey are summarized in Table~\ref{annualtable}. The
ESSENCE search is successful and our program finds roughly twice as
many objects with SN-like light curves than we can follow up
spectroscopically each year.


\begin{deluxetable}{ccc}
  \tablewidth{0pc} \tablecaption{Summary of the supernova yields from
    the first four years of ESSENCE observations.  } \tablehead{
    \colhead{Year} &    \colhead{Spectroscopically Confirmed Supernovae} & \colhead{Type Ia Supernovae} \\
  } \startdata
  2002    &      15     &     15 \\
  2003    &      37     &     33 \\
  2004    &      41     &     26 \\
  2005    &      46     &     28 \\
\enddata
\label{annualtable}
\end{deluxetable}

\pagebreak
\subsection {Image Analysis Pipeline}
\label{sec:pipeline}

The ESSENCE program requires immediate reduction of each night's data
(typically $\sim4$ GB each night), so it is more convenient to base
operations at the NOAO/CTIO offices in the nearby city of La Serena,
rather than directly at the telescope. Therefore, ESSENCE team members
carry out observations remotely from a terminal at La Serena,
communicating with telescope operators on the mountain via a
video-conferencing link.  Incoming data may be viewed by connecting
directly to computers at the telescope, which allows real-time quality
control, while in parallel the data are immediately transferred to
computing hardware in La Serena via an internet link.  

The analysis of ground-based imaging data is a complicated multi-stage
procedure, involving the removal of instrumental artifacts, calibration
of the data and measurements of the fluxes from the objects of interest.
The particular demands of a supernova survey place more demanding constraints
on the image analysis software.

First, the objects of interest are transient and appear in the data
masked by the background flux from their host galaxy. Past experience
has shown that the most reliable way to find these objects is via
image subtraction \citep{norgarrd89, perlmutter95, schmidt98}. For
each new image, an archival {}``template'' frame from a previous epoch
is subtracted pixel-by-pixel to remove constant sources, such as
galaxies, to reveal the supernovae.  Image subtraction software is not
part of standard analysis packages and we have invested significant
effort in developing robust and reliable methods necessary for our
project.

Second, supernovae must be detected in \emph{real time}. While it is a
part of our search strategy to revisit each field and build up a time
series of photometric measurements of all objects, we rely on
follow-up spectroscopic observations to verify the identity of
candidate transients as type Ia supernovae and to establish their
redshifts.  Because supernovae at the distances that give cosmological
leverage are faint ($m\sim22$) even at maximum light, it is preferable
to observe them near maximum light.  Type Ia supernovae rise to
maximum light roughly 20 days after explosion in their rest frame
\citep{riess-rise99,conley-rise06,garg-rise06}, and while time
dilation stretches the rise of a supernova by a factor of $1+z$, a
prompt detection allows us to schedule the spectroscopic observations
into the available time.  This real-time component adds a significant
demand on the analysis of the survey data: the data must be processed
automatically and reliably, in bulk, each night of the survey.

Finally, supernovae are rare events. We expect roughly one supernova
per MOSAIC field per month. Each MOSAIC field consists of
$4096\,\times\,2048\,\times\,8=67,000,000$ pixels, and we must be able
to reliably determine the $\sim$dozen pixels among those that contain
signal, often only marginally above background noise, from a
\emph{bona fide} type Ia supernova. \footnote{If there are roughly 10
needles to a gram and a typical haystack weighs 1000 kilograms, then
finding the part per $10^7$ supernova pixels in one frame is truly
like finding a needle in a haystack.  And we need to sift 20 per
night!}

We have developed a data pipeline that meets these demands, accepting
raw images directly from the telescope and automatically producing
lists of candidate objects only hours later. Floating point operations
are carried out by a variety of programs, either drawn from publicly
available astronomical software packages, such as IRAF\footnote{IRAF
is distributed by the National Optical Astronomy Observatory, which is
operated by AURA under cooperative agreement with the NSF.} , or
written by us (generally in C). These are tied together by a suite of
Perl scripts, which handle process management and bookkeeping.
Functionally, there are two separate piplines. The first of these
({}``mscpipe'') performs tasks relevant for full MOSAIC images and as
output divides each single MOSAIC field into sixteen 1k x 4k pixel
images corresponding to each CCD amplifier. From this point onward,
the {}``amplifier-images'' are processed through {}``photpipe'' and
each amplifier is effectively treated as an independent detector. 
We will refer to a single MOSAIC exposure as a MOSAIC field and the
subdivided images as subfields.  Below we provide a brief description
of the data processing, focusing in particular on those stages that
alter the data in ways significant for the analysis.

\subsubsection {Crosstalk correction}

Pairs of CCDs in the MOSAIC II imager are read out through single
electronics controllers, which, for some combinations of CCDs, results
in low-level cross-talk between the signals from different chips.  The
resulting effect is the appearance of {}``ghosts'' in one subfield of
bright objects appearing in another subfield. Fortunately, this effect
is small in magnitude, on the order of 0.1\%, and deterministic.  The
first stage of the \emph{mscpipe} pipeline uses the most recent values
of these cross-talk coefficients measured by the observatory staff and
subtracts these electronic artifacts from the affected portions of the
MOSAIC field, using the \emph{xtalk} task from the \emph{mscred}
package for IRAF.

\subsubsection {Astrometric calibration}

The transformation from pixel to sky coordinates is dominated by
distortions due to the optical system of the telescope that change
only slightly over long periods of time and generally take the form of
a polynomial in radius. Once the terms of this distortion function are
known, the astrometric calibration of any particular image reduces to
determining accurately the center of the distortion in that field,
essentially an offset in x and y and a rotation. This is accomplished
via the IRAF task \emph{msccmatch} from the \emph{mscred} package,
which matches objects in the image to an existing catalog of the field
with precise astrometry.  The current standard for astrometry is the
USNO CCD Astrograph Catalog 2 (UCAC, \citet{ucac2}) , which covers all
fields observed by ESSENCE.  However, since ESSENCE is a significantly
deeper survey, the Sloan Digital Sky Survey (SDSS,
\citet{york-sdss00}) provides a better photometric overlap.  We use
the SDSS (which itself is tied to UCAC) in the fields for which SDSS
has imaging data ($\sim73\%$ of ESSENCE fields), and default to UCAC
when there is no SDSS data.

When the supernovae are faint, their location in an image is poorly
constrained, and we must rely on the astrometric solution to tell us
precisely where to measure the flux.  Errors in positioning the PSF
produce an underestimate the object's flux.  Therefore, accurate
\emph{relative} astrometric calibration is essential to measuring
supernova flux at low signal-to-noise, since what matters is that we
are able to map pixels from invididual images to some consistent
coordinate system.  To this end, we generate astrometric catalogs from
our own data, which are themselves calibrated to either SDSS or UCAC.
All subsequent ESSENCE images are then registered to these internally
generated catalogs.

The astrometric solution is also used to ``warp'' each image to a
common pixel coordinate system, so that reference images can be
subtracted from them.  This is accomplished using the SWarp \citep{terapix}
software package, using a Lanczos windowed sinc function
to resample the pixels onto the new coordinate system.

\subsubsection {Flatfielding }

In order to obtain consistent flux measurements across the plane of
MOSAIC imager, we must normalize the response of all the pixels. This
flat fielding is achieved in three steps.  First, at the beginning of
each night, a screen inside the telescope dome is illuminated and
observed with the MOSAIC.  These high signal-to-noise flatfields
enable us to accurately correct for pixel-to-pixel variations and
other imperfections in the optical system, but introduces large-scale
variations (e.g. gradients due to non-uniform illumination of the
flat-field screen).  The second step is to combine all of the data
from a night's observations.  By masking all astronomical sources and
combining with a median statistic, an image of the illumination of the
focal plane due to the night sky is created.  This ``illumination
correction'' is also applied to the data, removing gradients of
$\sim1\%$ across a CCD.  Finally, we use the average difference in sky
level between each ccd to further regularize the overall flux scaling,
a $1\%$ correction to the dome flats.

\subsubsection {Photometric calibration}

Flat-fielded and SWarped images are then analyzed with the DoPHOT
photometry package \citep{schechter93} to identify and measure sources.  This
instrumental photometry is then calibrated against a catalog of objects
with known magnitudes, to determine the photometric zeropoint for the image.
Further discussion of photometric calibrations follows in Section 3.

\subsubsection {\label{imagesub}Image subtraction}

Each image is then differenced against a reference image.  This
suppresses all constant sources of flux and reveals transients such as
new supernovae.  To subtract two images taken under different
atmospheric conditions on different nights, we must correct for seeing
variations.  Our image subtraction software uses the algorithm devised
by Alard and Lupton \citep{alard1998,alard2000} to determine and apply
a convolution that matches the point-spread functions of the two
images prior to subtraction.  Improvements to the basic method have
produced a process that automatically, robustly, and reliably produces
clean subtractions in our data.

\subsubsection {Difference image object detection}

Object detection in the subtracted images is done with a modified
version of DOPHOT.  Resampling and convolution of the images
correlates flux between pixels, so we have modified the image
registration and subtraction software to propagate noise maps that
track these correlations.  These are then used to evaluate the
significance of objects detected in the difference image.

\subsection {Candidate selection}

Each observation of a single ESSENCE field yields hundreds of objects
detected above some significance threshold in the subtracted images.
These must be culled to produce a small set of objects that are
very likely to be type Ia supernovae and merit spectroscopic
observation on large telescopes.  We first apply a series of
software cuts, which include

\begin{itemize}

\item requiring that the object has the same PSF as stars in the
  original, unsubtracted image,
\item vetoing detections with significant amounts of pixels with negative flux,
  to guard against subtraction residuals, such as dipoles resulting
  from slight image misalignment,
\item vetoing variable sources identified in previous data 
  (variable stars, active galactic nuclei) and
\item requiring coincident detections in more than one passband or on
  subsequent nights, to reject asteroids (typically, two detections at 
       signal-to-noise ratio $> 5$ within a five day window).

\end{itemize}

While the above rules eliminate many of the false positives, we
ultimately rely on human inspection to reject the small fraction of
contaminants that evade these filters.  Common problems include
insufficient masking of pixels from bright stars, subtraction
artifacts, and variable objects that have not varied significantly in
previous ESSENCE data.

We also perform light curve fits to assess whether each object is
consistent with the known behavior of \snia. Preliminary fits of the
initial R and I photometry are compared with light curve templates of
a type Ia SN at z=0 in B and V filters (which are a good match for SN
Ia at z$\sim$0.4-0.5). The template light curve is representative of a
normal type Ia SN with $\Delta_{m15}=1.1$ mag, or stretch=1, and was
contructed from well-sampled light curves of low-z supernovae
\citep{prieto06}.  Using a chi-squared minimization, we determine the
the best fit values for time of B maximum, observed R and I magnitudes
at maximum, and stretch. We chose to use stretch here because it
parametrizes in a simple way the variety of light curve shapes of
\sneia\ \citep{goldhaber-stretch}.  Using the R and I magnitudes at
maximum and the stretch obtained from the fit, we can now estimate a
photometric redshift assuming that the candidate is a type Ia SN. A
standard $\Lambda$CDM with \OM$=0.3$, \OL$=0.7$ is used and no host
galaxy reddening is considered in these fits.

A summary of the data for each candidate object is presented on a web
page for human inspection.  We reject detections resulting from
subtraction artifacts by looking at image ``stamps'' at the position
of the supernova. The light curves from the preliminary photometry
enable us to reject objects that clearly have the wrong light curve
shape, color, and brightness for a supernova in the estimated redshift
range.

Because our spectroscopy resources are limited, we have to make
choices to observe the most promising targets.  
We select against objects right at the centers of galaxies
both because past experience has shown that these are frequently
active galactic nuclei and because contamination from the galaxy often
makes it impossible to positively identify the supernova in a
spectrum.  To avoid these problems, we select against candidates that
are superposed on point-like sources in the central pixel (0.27") of
the template image.  We know that the SN Ia in galaxy centers have a
broader distribution in apparent luminosity from SN Ia generally
\citep{jha-mlcs2k2}, but we do not expect any significant cosmological
bias from this selection criterion.

The objects that pass the above selection procedure are then sent to
team members for spectroscopic observation.  Because spectroscopy time
is limited and scheduled in advance, we are forced to prioritize those
objects that look most promising based on the data available at the
time.  Our survey is spectroscopy-limited: at the end of each
observing campaign, many objects remain that have Ia-like light
curves, but for which we were unable to obtain follow-up spectroscopy.
Nevertheless, we successfully detect and confirm new supernovae at a
rate of roughly one new object per night of 4m observing.

\pagebreak

\section{Spectroscopy}
\label{sec:spectroscopy}

\subsection{Observations}

Follow-up spectroscopic observations of ESSENCE targets are performed
at a wide variety of ground-based telescopes: the 10-m Keck I (+LRIS;
\citealt{oke95}) and II (+ESI, \citealt{sheinis02}; +DEIMOS,
\citealt{faber03}) telescopes; the 8-m VLT (+FORS1;
\citealt{appenzeller98}), Gemini North and South (+GMOS;
\citealt{hook03}) telescopes; the 6.5-m Magellan Baade (+IMACS;
\citealt{dressler04}) and Clay (+LDSS2; \citealt{mulchaey01}), MMT
(+BlueChannel; \citealt{schmidt89}) telescopes.  One target
(d100.waa7\_16; see \citealt{matheson05}) was confirmed as a Type Ia
supernova using the FAST spectrograph \citep{fabricant98} on the 1.5-m
Tillinghast telescope at the F.~L.~Whipple Observatory (FLWO). The
useful sample of supernovae from the ESSENCE program is limited by our
ability to identify \sneia\ spectroscopically.

Standard CCD processing and spectrum extraction are done with standard
IRAF routines. Except for the VLT data, all the spectra are extracted
using the optimal algorithm of \citet{horne86}. For the VLT data, we
apply a novel extraction method based on two-channel Richardson-Lucy
restoration \citep{blondin05} to minimize contamination of the
supernova spectrum by underlying galaxy light. The spectra are
wavelength calibrated using calibration-lamp spectra (usually HeNeAr).
For the flux calibration we use both standard IRAF routines and our
own IDL procedures, which include the removal of telluric lines using
the well-exposed continua of the spectrophotometric standard stars
\citep{wade88,matheson00b}.

\subsection{Supernova classification and redshift determination}

Supernovae are classified according to their early-time spectra
\citep[see][for a review]{filippenko97}. The distinctive spectroscopic
signature of a Type Ia supernova near maximum light is a deep
absorption feature due to Si~{\sc ii} $\lambda$6355, blueshifted by
$\sim 10000$~km s$^{-1}$.  Their spectra are further characterized by
the absence of hydrogen and helium lines, although hydrogen has been
detected in the spectrum of the Type Ia supernova SN~2002ic
\citep{hamuy03,wood-vasey-2002ic} (\citet{benetti-2002ic} classify
this object as an Type Ib/c). Spectra of Type Ib supernovae are
characterized by a weaker Si~{\sc ii} $\lambda$6355 absorption, and by
the presence of lines of He~I.  Spectra of Type Ic supernovae are
devoid of He~I lines and display only a weak Si~{\sc ii} $\lambda$6355
absorption. Thus, in principle, SNe~Ib/c are readily distinguishable
from \sneia.

At high ($z \gtrsim 0.4$) redshifts, however, the defining Si~{\sc ii}
$\lambda$6355 feature in \sneia\ is redshifted out of the optical range
of most of the spectrographs we use, so features blueward of this must
be used to establish the type. The most prominent of these, the
Ca~{\sc ii} H\&K $\lambda\lambda$3934,3968 doublet, is also present in
SNe~Ib/c and does not discriminate between the various supernova
types.  Instead, the identification of \sneia\ relies on
weaker features (e.g., Si~{\sc ii} $\lambda4130$, Mg~{\sc ii}
$\lambda4481$, Fe~{\sc ii} $\lambda4555$, Si~{\sc iii} $\lambda4560$,
S~{\sc ii} $\lambda4816$, and Si~{\sc ii} $\lambda5051$).

While the above gives the general defining features of Ia spectra, in
practice, identfying \sneia\ can be difficult in low signal-to-noise
spectra, particular when trying to discriminate between \sneia\ and
SNe~Ib/c.  In addition, we would like to establish objective and
reproducible criteria for classifying objects, rather than relying on
subjective assessments of noisy data.  Therefore, we have developed an
algorithm (SuperNova IDentification, or SNID; \citealt{blondin-snid})
also used by \citet{matheson05}, which we use here to establish our
final \snia\ sample. This algorithm cross-correlates an input spectrum
with a library of supernova spectra, without attempting to directly
identify specific features, and a redshift is determined based on the
shift in wavelength that maximizes the correlation.  The spectral
database currently spans all supernova types and covers a wide range
of ages, containing 796 spectra of 64 \sneia\ (including spectra of
1991T-like and 1991bg-like objects), 288 spectra of 17 SNe~Ib/c, and
353 spectra of 10 SNe~II. We also include spectra of galaxies, AGNs,
and stars to identify spectra that are not consistent with a supernova
\citep[see also][]{matheson05}. The results of the SNID analysis are
shown in Table \ref{snid_table}.

The correlation redshift is valid when templates of the correct
supernova type are used.  We also use SNID to determine the supernova
type, by computing the absolute fraction of ``good'' correlations that
correspond to supernovae of different types. The supernova
types/subtypes in the SNID spectral database are: Ia/Ia-norm, Ia-pec,
Ia-91t, Ia-91bg; Ib/Ib-norm, Ib-pec, IIb; Ic/Ic-norm, Ic-pec,
Ic-broad; II/II-norm, II-pec, IIL, IIn, IIP, IIb. ``Norm'' and ``pec''
subtypes are used to identify the spectroscopically ``normal'' and
``peculiar'' supernovae of a given type, respectively. For type Ia
supernovae, ``91t'' and ``91bg'' indicates spectra that resemble those
of the overluminous SN~1991T and the underluminous SN~1991bg,
respectively. The spectra that correspond to the ``Ia-pec'' category
in this case are those of SNe~2000cx \citep{li-2000cx,candia-2000cx}
and 2002cx \citep{li-2002cx}. For type Ic supernovae, `Ic-broad'' is
used to identify broad-lined SNe~Ic, (often referred to as
``hypernovae'' in the literature), some of which are associated with
Gamma-Ray Bursts.  The notation used for the type II subtypes are
commonly used in the literature. Note that type IIb supernovae (whose
spectra evolve from a type II to a type Ib, as, e.g., in SN 1993J--
see \citealt{matheson00a}) are included both in the ``Ib'' and ``II''
types.

If the redshift of the supernova host galaxy can be measured using
narrow emission or absorption lines, we force SNID to look for
correlations at the galaxy redshift ($\pm 0.03$) to determine the
supernova type/subtype; otherwise the redshift is left as a free
parameter.  We assert a supernova to be of a given type (i.e., Ia, Ib,
Ic, II, see Table \ref{snid_table}, column 3) when the absolute
fraction of ``good'' correlations that correspond to this type exceeds
50\%. In addition, we require the best-match supernova template to be
of the same type. We determine the supernova subtype by requiring that
the absolute fraction of ``good'' correlations that correspond to this
subtype exceeds 50\%, {\it and} that it corresponds to the
previously-determined type. We also require that the best-match
supernova template is of the same subtype.

The requirement that an object must have a correlation fraction above
50\% is motivated by the desire to have a quantative figure of merit
that determines when the spectral information is strong enough to make
a positive identification.  Out of all the spectra that were
considered to be those of possible supernovae, 28 did not meet the
above criteria for a positive classification (see Table
\ref{snid_table}). Assessing the likelihood that a spectrum matches
that of particular known object more closely than others is a
challenging statistical problem, especially in the presence of
intrinsic and only partially understood variance in the populations of
supernovae.  See \citet{blondin-snid} for a detailed discussion of
ongoing work to better understand these issues.

\begin{figure}
  \plotone{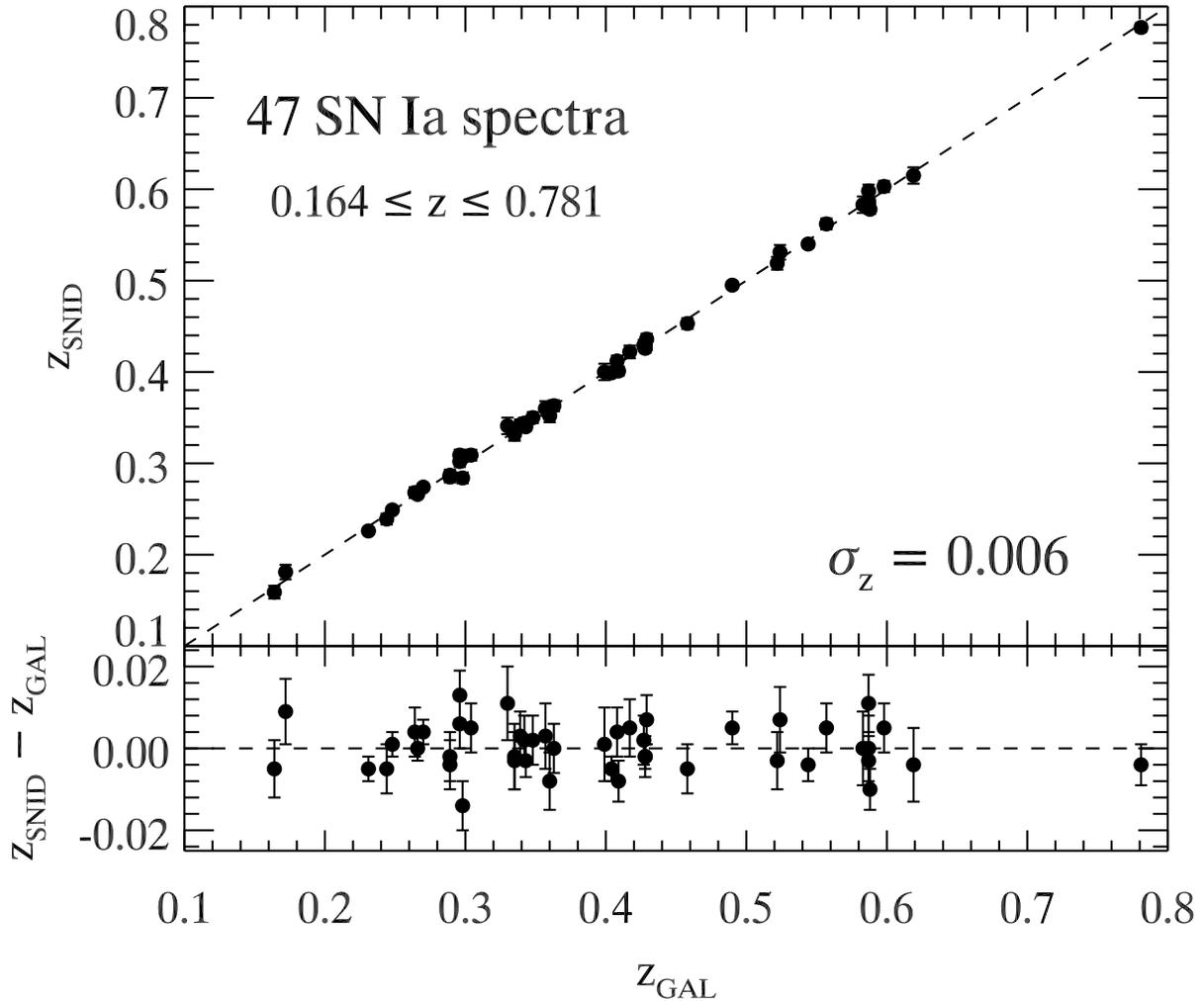}
  \caption{Comparison of ESSENCE SN~Ia redshifts obtained from narrow
    emission and/or absorption lines in the host galaxy spectrum
    ($z_{\rm GAL}$), and from cross-correlations with a library of SN~Ia
    spectral templates ($z_{\rm SN}$). The correspondence is excellent,
    with a standard deviation from the one-to-one correspondence of only
    $\sim 0.006$ \citep[see also][]{matheson05}.  Only the 47 ESSENCE
    supernovae for which it was possible to measure host galaxy redshifts
    are used.
    \label{snidfig}}
\end{figure}

\begin{figure}
\plotone{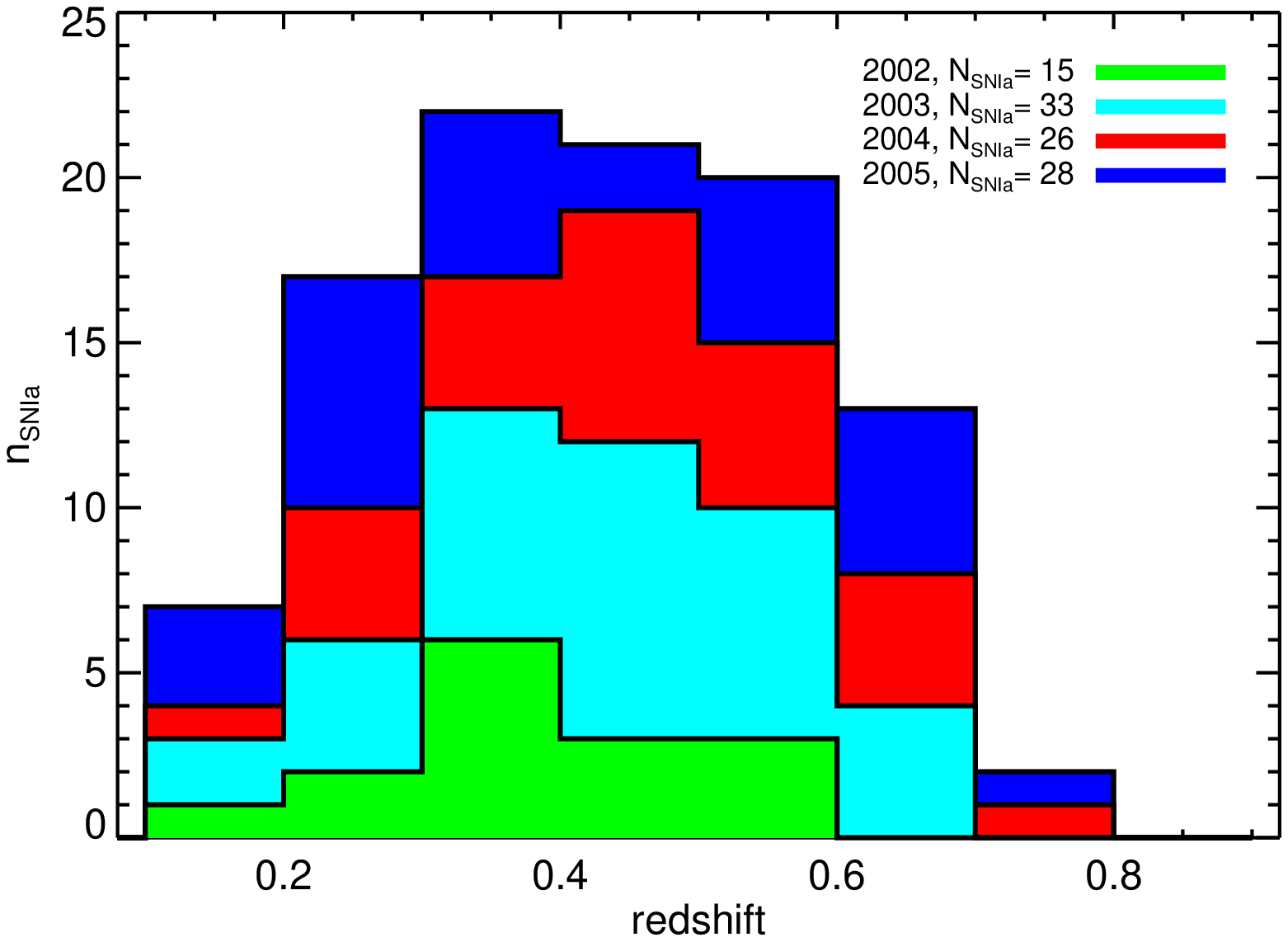}
  \caption{Redshift histogram of spectroscopically confirmed ESSENCE
    \snia\ (all objects whose type Ia correlations exceed 50\%).
    \label{redshift_hist}}
\end{figure}

The redshift is then determined from the supernova spectrum alone in a
second SNID run by considering correlations with templates of the
determined type and subtype. No {\it a priori} information on redshift
is used in this second run.  The supernova redshift is reported as the
median redshift of all ``good'' correlations, and the redshift error
as the standard deviation of these same redshifts. When there is only
one ``good'' correlation for an input spectrum (objects d087, h311,
and p524 in Table~\ref{snid_table}), we quote the redshift as that of
the best-match template and the associated error as the formal
redshift error for that template (see \citealt{blondin-snid}). We only
report a SN redshift when a secure type is determined.  In
\citet{matheson05} we found an excellent agreement between the SNID
correlation redshift and the redshift of the supernova host galaxy
when it is known from other methods. Figure \ref{snidfig} again shows
that the SNID redshifts agree well with the galaxy redshifts, with a
typical uncertainty $\lesssim 0.01$ in the redshift range $[0.1-0.8]$.
Figure \ref{redshift_hist} shows the redshift distribution of the
spectroscopically confirmed \sneia\ from the first four years of
ESSENCE.

\begin{deluxetable}{lcccllrrrrccc}
\rotate
\tabletypesize{\scriptsize}
\tablewidth{0pc}
\tablecaption{Types and Redshifts of ESSENCE Supernovae.
} 
\tablehead{
  \colhead{IAUC ID} & \colhead{RA [J2000]} & \colhead{Dec [J2000]} & \colhead{ESSENCE ID} & \colhead{Type}       & \colhead{Subtype}    & \colhead{\%Subtype} & \colhead{\%Ia}     & \colhead{\%Ib/c}   & \colhead{\%II}     & \colhead{$z_{\rm GAL}$} & \colhead{$z_{\rm SNID}$} & \colhead{$\sigma_z$} \\
}
\startdata
2002iu                    & 00:13:33.10  & -10:13:09.92 & b003       & Ia   & Ia-norm    &     74.2 &    100.0 &      0.0 &      0.0 &      --- &    0.115 &    0.006 \\ 
2002iv                    & 02:19:16.11  & -07:44:06.72 & b004       & Ia   & Ia-91t     &     64.4 &     98.3 &      1.7 &      0.0 &    0.231 &    0.226 &    0.003 \\ 
2002jq                    & 23:35:57.96  & -10:05:56.88 & b008       & Ia   & Ia-norm    &     65.1 &     81.4 &     11.6 &      7.0 &      --- &    0.474 &    0.004 \\ 
2002iy                    & 02:30:40.00  & -08:11:40.50 & b010       & Ia   & Ia-norm    &     73.6 &     82.4 &     17.6 &      0.0 &    0.587 &    0.590 &    0.006 \\ 
2002iz                    & 02:31:20.73  & -08:36:13.12 & b013       & Ia   & Ia-norm    &     85.8 &     98.6 &      1.4 &      0.0 &    0.428 &    0.426 &    0.004 \\ 
2002ja                    & 23:30:09.66  & -09:35:01.75 & b016       & Ia   & Ia-norm    &    100.0 &    100.0 &      0.0 &      0.0 &      --- &    0.329 &    0.003 \\ 
2002jb                    & 23:29:44.14  & -09:36:34.25 & b017       & Ia   & Ia-norm    &     75.5 &    100.0 &      0.0 &      0.0 &      --- &    0.258 &    0.007 \\ 
2002jr                    & 02:04:41.03  & -05:09:40.73 & b020       & Ia   & Ia-norm    &     81.8 &    100.0 &      0.0 &      0.0 &      --- &    0.425 &    0.003 \\ 
2002jc                    & 02:07:27.28  & -03:50:20.73 & b022       & Ia   & Ia-norm    &     55.7 &     65.7 &     24.3 &     10.0 &      --- &    0.540 &    0.008 \\ 
2002js                    & 02:20:35.39  & -09:34:43.90 & b023       & Ia   & Ia-norm    &     90.9 &    100.0 &      0.0 &      0.0 &      --- &    0.550 &    0.007 \\ 
2002jd                    & 00:28:38.39  & +00:40:29.29 & b027       & Ia   & Ia-norm    &     79.2 &     96.6 &      3.4 &      0.0 &      --- &    0.318 &    0.005 \\ 
2002jt                    & 00:13:36.70  & -10:08:24.00 & c003       & Ia   & ---{\tablenotemark{a}}&--- &100.0&      0.0 &      0.0 &      --- &    0.382 &    0.002 \\ 
2002ju                    & 02:20:11.00  & -09:04:37.50 & c012       & Ia   & Ia-norm    &     72.6 &    100.0 &      0.0 &      0.0 &    0.348 &    0.350 &    0.006 \\ 
2002jw                    & 02:30:00.52  & -08:36:22.41 & c015       & Ia   & Ia-norm    &     76.6 &    100.0 &      0.0 &      0.0 &    0.357 &    0.362 &    0.008 \\ 
---                       & 00:28:03.16  & +00:37:50.43 & c023       & Ia   & Ia-norm    &    100.0 &    100.0 &      0.0 &      0.0 &    0.399 &    0.400 &    0.009 \\ 
2003jo                    & 23:25:24.03  & -09:26:00.63 & d033       & Ia   & Ia-norm    &     76.0 &     96.0 &      4.0 &      0.0 &    0.524 &    0.531 &    0.008 \\ 
2003jj                    & 01:07:58.52  & +00:03:01.89 & d058       & Ia   & Ia-norm    &     85.0 &     95.0 &      5.0 &      0.0 &    0.583 &    0.583 &    0.009 \\ 
2003jn                    & 02:29:21.21  & -09:02:15.57 & d083       & Ia   & Ia-91t     &     56.5 &    100.0 &      0.0 &      0.0 &      --- &    0.333 &    0.002 \\ 
2003jm                    & 02:28:50.93  & -09:09:58.14 & d084       & Ia   & Ia-norm    &     68.6 &    100.0 &      0.0 &      0.0 &    0.522 &    0.519 &    0.007 \\ 
2003jv                    & 23:27:58.22  & -08:57:11.82 & d085       & Ia   & Ia-91t     &    100.0 &    100.0 &      0.0 &      0.0 &    0.405 &    0.401 &    0.001 \\ 
2003ju                    & 23:27:01.71  & -09:24:04.49 & d086       & Ia   & Ia-norm    &     87.5 &    100.0 &      0.0 &      0.0 &      --- &    0.205 &    0.003 \\ 
2003jr                    & 01:11:06.23  & +00:13:44.21 & d087       & Ia   & Ia-norm    &    100.0 &    100.0 &      0.0 &      0.0 &    0.340 &    0.337{\tablenotemark{b}} &    0.009 \\ 
2003jl                    & 02:28:28.56  & -08:08:44.74 & d089       & Ia   & Ia-norm    &     92.3 &    100.0 &      0.0 &      0.0 &    0.429 &    0.436 &    0.006 \\ 
2003js                    & 02:29:52.15  & -08:32:28.09 & d093       & Ia   & Ia-91t     &     63.1 &     93.4 &      6.6 &      0.0 &    0.363 &    0.360 &    0.004 \\ 
2003jt                    & 02:31:54.60  & -08:35:48.43 & d097       & Ia   & Ia-norm    &     95.7 &    100.0 &      0.0 &      0.0 &      --- &    0.436 &    0.008 \\ 
2003ji                    & 02:07:54.84  & -03:28:28.40 & d099       & Ia   & Ia-norm    &     77.5 &     96.9 &      2.0 &      1.0 &      --- &    0.211 &    0.003 \\ 
2003jq                    & 23:30:51.19  & -09:28:33.95 & d100       & Ia   & Ia-norm    &     67.8 &     98.3 &      1.7 &      0.0 &      --- &    0.156 &    0.003 \\ 
2003jw                    & 02:31:06.84  & -08:45:36.51 & d117       & Ia   & Ia-norm    &     84.6 &    100.0 &      0.0 &      0.0 &    0.296 &    0.309 &    0.006 \\ 
2003jy                    & 02:10:53.98  & -04:25:49.76 & d149       & Ia   & Ia-norm    &    100.0 &    100.0 &      0.0 &      0.0 &    0.339 &    0.342 &    0.006 \\ 
2003kk                    & 23:25:36.06  & -09:31:44.70 & e020       & Ia   & Ia-norm    &     88.4 &    100.0 &      0.0 &      0.0 &    0.164 &    0.159 &    0.007 \\ 
2003kl                    & 01:09:48.80  & +01:00:05.58 & e029       & Ia   & Ia-norm    &     74.7 &    100.0 &      0.0 &      0.0 &    0.335 &    0.332 &    0.008 \\ 
2003km                    & 02:30:01.00  & -09:04:35.89 & e108       & Ia   & Ia-norm    &    100.0 &    100.0 &      0.0 &      0.0 &      --- &    0.469 &    0.005 \\ 
2003kn                    & 02:09:15.55  & -03:35:41.38 & e132       & Ia   & Ia-norm    &     76.3 &    100.0 &      0.0 &      0.0 &    0.244 &    0.239 &    0.006 \\ 
2003ko                    & 02:11:06.48  & -03:47:56.09 & e136       & Ia   & Ia-norm    &     85.1 &     99.2 &      0.8 &      0.0 &    0.360 &    0.352 &    0.007 \\ 
2003kt                    & 02:33:47.01  & -08:36:22.09 & e138       & Ia   & Ia-norm    &    100.0 &    100.0 &      0.0 &      0.0 &      --- &    0.612 &    0.009 \\ 
2003kq                    & 02:31:04.09  & -08:10:56.64 & e140       & Ia   & Ia-norm    &    100.0 &    100.0 &      0.0 &      0.0 &    0.606 &    0.631 &    0.007 \\ 
2003kp                    & 02:31:02.64  & -08:39:50.81 & e147       & Ia   & Ia-norm    &    100.0 &    100.0 &      0.0 &      0.0 &      --- &    0.645 &    0.010 \\ 
2003kr                    & 02:31:20.96  & -08:36:14.16 & e148       & Ia   & Ia-norm    &    100.0 &    100.0 &      0.0 &      0.0 &    0.427 &    0.429 &    0.006 \\ 
2003ks                    & 02:31:34.54  & -08:36:46.41 & e149       & Ia   & Ia-norm    &     81.4 &     98.6 &      1.4 &      0.0 &      --- &    0.497 &    0.006 \\ 
2003ku{\tablenotemark{c}} & 01:08:36.25  & -00:33:20.78 & e315       & ---  & ---        &      --- &      --- &      --- &      --- &      --- &      --- &      --- \\ 
2003kv{\tablenotemark{c}} & 02:09:42.52  & -03:46:48.58 & e531       & ---  & ---        &      --- &      --- &      --- &      --- &      --- &      --- &      --- \\ 
2003lh                    & 02:10:19.51  & -04:59:32.30 & f011       & Ia   & Ia-norm    &    100.0 &    100.0 &      0.0 &      0.0 &      --- &    0.539 &    0.004 \\ 
2003le                    & 01:08:08.73  & +00:27:09.74 & f041       & Ia   & Ia-norm    &     68.8 &    100.0 &      0.0 &      0.0 &      --- &    0.561 &    0.006 \\ 
2003lf                    & 01:08:49.81  & -00:44:13.49 & f076       & Ia   & Ia-norm    &     82.2 &    100.0 &      0.0 &      0.0 &      --- &    0.410 &    0.007 \\ 
2003lm                    & 23:24:25.51  & -08:45:51.11 & f096       & Ia   & Ia-norm    &     88.5 &    100.0 &      0.0 &      0.0 &    0.408 &    0.412 &    0.006 \\ 
2003ll                    & 02:35:41.19  & -08:06:29.55 & f216       & Ia   & Ia-norm    &     75.0 &    100.0 &      0.0 &      0.0 &    0.596 &    0.599 &    0.005 \\ 
2003lk{\tablenotemark{d}} & 02:11:12.82  & -04:13:52.11 & f221       & ---  & ---        &     --- &     33.3 &     66.7 &      0.0 &    0.442 &      --- &      --- \\ 
2003ln                    & 23:30:27.15  & -08:35:46.98 & f231       & Ia   & Ia-norm    &    100.0 &    100.0 &      0.0 &      0.0 &      --- &    0.619 &    0.008 \\ 
2003lj                    & 01:12:10.03  & +00:19:51.29 & f235       & Ia   & Ia-norm    &     87.8 &    100.0 &      0.0 &      0.0 &    0.417 &    0.422 &    0.007 \\ 
2003li                    & 02:27:47.29  & -07:33:46.16 & f244       & Ia   & Ia-norm    &    100.0 &    100.0 &      0.0 &      0.0 &    0.544 &    0.540 &    0.004 \\ 
---   {\tablenotemark{d}} & 02:27:26.51  & -08:42:24.88 & f301       & ---  & ---        &     50.0 &     75.0 &     14.3 &     10.7 &      --- &      --- &      --- \\ 
---                       & 02:29:22.39  & -08:37:38.38 & f308       & Ia   & Ia-norm    &     66.7 &    100.0 &      0.0 &      0.0 &      --- &    0.394 &    0.009 \\ 
2004fi{\tablenotemark{c}} & 23:29:45.35  & -08:54:36.34 & g001       & ---  & ---        &      --- &      --- &      --- &      --- &    0.265 &      --- &      --- \\ 
2004fh                    & 23:28:27.20  & -08:36:55.17 & g005       & Ia   & Ia-norm    &     72.9 &    100.0 &      0.0 &      0.0 &      --- &    0.218 &    0.007 \\ 
2004fj                    & 01:09:51.07  & +00:27:20.95 & g043       & II   & IIP        &    100.0 &      0.0 &      0.0 &    100.0 &    0.187 &    0.193 &    0.002 \\ 
2004fn                    & 23:30:20.12  & -09:58:30.67 & g050       & Ia   & Ia-norm    &    100.0 &    100.0 &      0.0 &      0.0 &    0.605 &    0.633 &    0.006 \\ 
2004fm                    & 23:26:58.14  & -09:37:19.45 & g052       & Ia   & Ia-norm    &     80.0 &    100.0 &      0.0 &      0.0 &      --- &    0.383 &    0.008 \\ 
2004fl{\tablenotemark{c}} & 23:26:57.92  & -09:37:19.11 & g053       & ---  & ---        &      --- &      --- &      --- &      --- &      --- &      --- &      --- \\ 
2004fk                    & 01:13:35.84  & -00:09:27.56 & g055       & Ia   & Ia-norm    &     79.3 &    100.0 &      0.0 &      0.0 &    0.296 &    0.302 &    0.006 \\ 
---                       & 23:27:37.16  & -09:35:20.96 & g097       & Ia   & Ia-norm    &     62.8 &     81.4 &     18.6 &      0.0 &    0.343 &    0.340 &    0.004 \\ 
2004fo                    & 01:13:28.97  & +00:35:16.26 & g120       & Ia   & Ia-norm    &     94.7 &    100.0 &      0.0 &      0.0 &      --- &    0.510 &    0.009 \\ 
---                       & 02:09:49.63  & -04:10:55.07 & g133       & Ia   & Ia-norm    &     75.0 &     98.8 &      0.0 &      1.2 &      --- &    0.421 &    0.003 \\ 
---                       & 23:28:37.70  & -08:45:04.01 & g142       & Ia   & Ia-norm    &     58.2 &     98.5 &      1.5 &      0.0 &    0.404 &    0.399 &    0.003 \\ 
2004fq{\tablenotemark{c}} & 23:27:45.64  & -08:31:12.77 & g151       & ---  & ---        &      --- &      --- &      --- &      --- &    0.146 &      --- &      --- \\ 
2004fs                    & 02:31:19.95  & -08:49:21.67 & g160       & Ia   & Ia-norm    &     89.5 &    100.0 &      0.0 &      0.0 &      --- &    0.493 &    0.003 \\ 
2004fr{\tablenotemark{c}} & 02:28:43.77  & -08:54:24.05 & g166       & ---  & ---        &      --- &      --- &      --- &      --- &    0.202 &      --- &      --- \\ 
2004ft{\tablenotemark{c}} & 02:33:32.63  & -08:09:34.10 & g199       & ---  & ---        &      --- &      --- &      --- &      --- &      --- &      --- &      --- \\ 
---   {\tablenotemark{c}} & 23:27:15.69  & -09:27:59.76 & g225       & ---  & ---        &      --- &      --- &      --- &      --- &      --- &      --- &      --- \\ 
---   {\tablenotemark{d}} & 01:11:56.31  & +00:07:27.71 & g230       & ---  & ---        &      --- &     50.0 &      0.0 &     50.0 &    0.392 &      --- &      --- \\ 
---                       & 23:30:41.83  & -08:34:10.98 & g240       & Ia   & Ia-norm    &     86.7 &    100.0 &      0.0 &      0.0 &      --- &    0.687 &    0.005 \\ 
---   {\tablenotemark{c}} & 02:04:27.01  & -03:35:43.72 & g276       & ---  & ---        &      --- &      --- &      --- &      --- &    0.244 &      --- &      --- \\ 
2004ha                    & 02:04:27.01  & -04:52:46.03 & h283       & Ia   & Ia-norm    &     85.7 &    100.0 &      0.0 &      0.0 &      --- &    0.502 &    0.008 \\ 
---                       & 02:31:40.67  & -08:49:03.35 & h300       & Ia   & Ia-norm    &    100.0 &    100.0 &      0.0 &      0.0 &      --- &    0.687 &    0.012 \\ 
2004hc                    & 23:24:32.67  & -08:41:03.55 & h311       & Ia   & Ia-norm    &    100.0 &    100.0 &      0.0 &      0.0 &      --- &    0.741{\tablenotemark{b}} &     0.011 \\ 
2004hd                    & 02:08:48.21  & -04:26:10.42 & h319       & Ia   & Ia-norm    &    100.0 &    100.0 &      0.0 &      0.0 &    0.490 &    0.495 &    0.004 \\ 
2004he                    & 02:29:48.79  & -08:20:45.94 & h323       & Ia   & Ia-norm    &    100.0 &    100.0 &      0.0 &      0.0 &    0.598 &    0.603 &    0.006 \\ 
2004hf                    & 02:32:00.14  & -08:42:23.89 & h342       & Ia   & Ia-norm    &    100.0 &    100.0 &      0.0 &      0.0 &      --- &    0.421 &    0.002 \\ 
2004hg{\tablenotemark{c}} & 02:34:55.19  & -08:30:43.64 & h345       & ---  & ---        &      --- &      --- &      --- &      --- &      --- &      --- &      --- \\ 
2004hi                    & 02:08:38.84  & -05:08:11.79 & h359       & Ia   & Ia-norm    &     46.8 &     68.1 &     31.9 &      0.0 &      --- &    0.348 &    0.004 \\ 
2004hh                    & 02:06:25.02  & -04:38:04.09 & h363       & Ia   & Ia-norm    &     69.0 &     97.7 &      0.0 &      2.3 &      --- &    0.213 &    0.006 \\ 
2004hj                    & 02:29:41.94  & -08:43:49.42 & h364       & Ia   & Ia-norm    &    100.0 &    100.0 &      0.0 &      0.0 &      --- &    0.344 &    0.007 \\ 
2004hk{\tablenotemark{c}} & 23:27:04.39  & -08:38:45.11 & k396       & ---  & ---        &      --- &      --- &      --- &      --- &      --- &      --- &      --- \\ 
---                       & 23:26:11.77  & -08:50:17.50 & k411       & Ia   & Ia-norm    &     78.6 &     85.7 &     14.3 &      0.0 &      --- &    0.564 &    0.006 \\ 
2004hl                    & 01:13:38.17  & -00:27:39.03 & k425       & Ia   & Ia-norm    &     82.9 &     97.1 &      0.0 &      2.9 &    0.270 &    0.274 &    0.003 \\ 
2004hm                    & 02:28:03.12  & -07:42:29.70 & k429       & Ia   & Ia-norm    &     66.7 &    100.0 &      0.0 &      0.0 &    0.172 &    0.181 &    0.008 \\ 
2004hn                    & 01:13:32.39  & +00:37:15.38 & k430       & Ia   & Ia-norm    &    100.0 &    100.0 &      0.0 &      0.0 &      --- &    0.582 &    0.010 \\ 
---   {\tablenotemark{c}} & 01:13:38.17  & -00:27:39.03 & k432       & ---  & ---        &      --- &      --- &      --- &      --- &      --- &      --- &      --- \\ 
2004hq                    & 02:30:18.04  & -08:22:25.01 & k441       & Ia   & Ia-norm    &     81.0 &    100.0 &      0.0 &      0.0 &      --- &    0.680 &    0.010 \\ 
2004hp{\tablenotemark{c}} & 02:09:35.52  & -03:46:23.53 & k443       & ---  & ---        &      --- &      --- &      --- &      --- &      --- &      --- &      --- \\ 
2004hr                    & 01:08:48.34  & +00:00:49.49 & k448       & Ia   & Ia-norm    &    100.0 &    100.0 &      0.0 &      0.0 &    0.409 &    0.401 &    0.005 \\ 
---   {\tablenotemark{c}} & 02:31:11.80  & -07:47:34.13 & k467       & ---  & ---        &      --- &      --- &      --- &      --- &      --- &      --- &      --- \\ 
2004hs                    & 02:09:33.69  & -04:13:03.93 & k485       & Ia   & Ia-norm    &     93.3 &    100.0 &      0.0 &      0.0 &      --- &    0.416 &    0.005 \\ 
---   {\tablenotemark{c}} & 02:30:24.32  & -07:53:20.95 & k490       & ---  & ---        &      --- &      --- &      --- &      --- &    0.715 &      --- &      --- \\ 
---   {\tablenotemark{c}} & 01:08:22.01  & -00:05:46.65 & m001       & ---  & ---        &      --- &      --- &      --- &      --- &      --- &      --- &      --- \\ 
---                       & 02:05:27.31  & -04:42:54.05 & m003       & II   & ---{\tablenotemark{a}} &34.2& 2.6 &      0.0 &     97.4 &      --- &    0.219 &    0.001 \\ 
---   {\tablenotemark{c}} & 02:30:27.27  & -09:16:10.23 & m006       & ---  & ---        &      --- &      --- &      --- &      --- &    0.057 &      --- &      --- \\ 
---                       & 02:31:46.24  & -09:16:25.65 & m010       & Ib   & Ib-norm    &    100.0 &      0.0 &    100.0 &      0.0 &    0.216 &    0.222 &    0.001 \\ 
---                       & 02:08:06.23  & -04:03:51.16 & m011       & II   & IIP        &     78.1 &      0.0 &      0.0 &    100.0 &    0.205 &    0.211 &    0.002 \\ 
---                       & 02:07:12.91  & -04:26:40.06 & m014       & II   & IIP        &     50.0 &      0.0 &      0.0 &    100.0 &    0.200 &    0.212 &    0.003 \\ 
---                       & 23:30:02.70  & -08:33:36.57 & m022       & Ia   & ---{\tablenotemark{a}}&---& 93.8 &      1.8 &      4.4 &      --- &    0.240 &    0.003 \\ 
---                       & 23:28:39.97  & -09:19:50.00 & m026       & Ia   & ---{\tablenotemark{a}}&---& 97.8 &      2.2 &      0.0 &    0.655 &    0.653 &    0.008 \\ 
---                       & 01:09:15.01  & +00:08:14.80 & m027       & Ia   & Ia-norm    &     72.2 &     92.6 &      7.4 &      0.0 &    0.289 &    0.286 &    0.006 \\ 
---                       & 23:29:35.34  & -09:58:46.33 & m032       & Ia   & Ia-norm    &     80.2 &     96.5 &      3.5 &      0.0 &      --- &    0.155 &    0.004 \\ 
---                       & 02:27:50.33  & -07:59:11.62 & m034       & Ia   & Ia-norm    &     96.3 &    100.0 &      0.0 &      0.0 &    0.557 &    0.562 &    0.006 \\ 
---                       & 02:05:10.83  & -04:47:13.94 & m038       & II   & IIP        &     94.4 &      5.6 &      0.0 &     94.4 &    0.051 &    0.054 &    0.003 \\ 
---                       & 02:28:04.63  & -07:42:44.29 & m039       & Ia   & Ia-norm    &     84.4 &    100.0 &      0.0 &      0.0 &    0.248 &    0.249 &    0.003 \\ 
---                       & 02:09:49.78  & -04:45:10.65 & m041       & II   & ---{\tablenotemark{a}}&---& 22.8 &      0.0 &     77.2 &      --- &    0.220 &    0.004 \\ 
---                       & 23:29:51.73  & -08:56:46.07 & m043       & Ia   & Ia-norm    &     57.3 &     99.5 &      0.0 &      0.5 &    0.266 &    0.266 &    0.003 \\ 
---                       & 02:10:56.77  & -04:27:29.90 & m057       & Ia   & ---{\tablenotemark{a}}&---& 95.5 &      0.4 &      4.1 &    0.180 &    0.184 &    0.003 \\ 
---                       & 01:09:52.90  & +00:36:19.03 & m062       & Ia   & ---{\tablenotemark{a}}&---&100.0 &      0.0 &      0.0 &    0.314 &    0.317 &    0.005 \\ 
---                       & 23:24:42.28  & -08:29:07.82 & m075       & Ia   & ---{\tablenotemark{a}}&---&100.0 &      0.0 &      0.0 &    0.100 &    0.102 &    0.001 \\ 
---                       & 01:08:56.35  & +00:39:25.38 & m138       & Ia   & Ia-norm    &     66.7 &    100.0 &      0.0 &      0.0 &    0.587 &    0.582 &    0.004 \\ 
---                       & 23:23:57.83  & -08:27:08.33 & m139       & II   & ---{\tablenotemark{a}}&---&  0.0 &      0.0 &    100.0 &    0.212 &      --- &      --- \\ 
---                       & 23:24:03.53  & -09:23:18.24 & m158       & Ia   & ---{\tablenotemark{a}}&---& 95.2 &      4.8 &      0.0 &      --- &    0.463 &    0.007 \\ 
---                       & 02:28:52.20  & -07:42:09.78 & m193       & Ia   & Ia-norm    &    100.0 &    100.0 &      0.0 &      0.0 &    0.330 &    0.341 &    0.009 \\ 
---                       & 02:06:03.69  & -04:39:59.12 & m226       & Ia   & ---{\tablenotemark{a}}&---& 95.2 &      4.8 &      0.0 &    0.675 &    0.671 &    0.004 \\ 
---   {\tablenotemark{c}} & 01:14:33.08  & -00:26:23.18 & n246       & ---  & ---        &      --- &      --- &      --- &      --- &    0.706 &      --- &      --- \\ 
---                       & 02:28:09.01  & -07:47:49.56 & n256       & Ia   & Ia-norm    &    100.0 &    100.0 &      0.0 &      0.0 &      --- &    0.631 &    0.012 \\ 
---                       & 02:06:42.35  & -04:22:37.01 & n258       & Ia   & Ia-norm    &     50.0 &     81.6 &     18.4 &      0.0 &      --- &    0.522 &    0.007 \\ 
---                       & 02:05:14.95  & -04:56:39.08 & n263       & Ia   & Ia-norm    &     79.9 &    100.0 &      0.0 &      0.0 &      --- &    0.368 &    0.007 \\ 
---                       & 01:13:06.51  & +00:30:04.86 & n271       & II   & IIP        &     85.2 &      0.0 &      0.0 &    100.0 &      --- &    0.241 &    0.004 \\ 
---                       & 23:28:17.55  & -09:23:12.38 & n278       & Ia   & Ia-norm    &     78.5 &    100.0 &      0.0 &      0.0 &    0.304 &    0.309 &    0.006 \\ 
---                       & 23:23:51.35  & -08:23:18.47 & n285       & Ia   & Ia-norm    &     64.5 &     81.4 &     14.5 &      4.1 &      --- &    0.528 &    0.006 \\ 
---   {\tablenotemark{c}} & 02:29:00.48  & -09:02:52.96 & n322       & ---  & ---        &      --- &      --- &      --- &      --- &      --- &      --- &      --- \\ 
---                       & 23:29:58.59  & -08:53:12.45 & n326       & Ia   & Ia-norm    &     79.8 &    100.0 &      0.0 &      0.0 &    0.264 &    0.268 &    0.006 \\ 
---                       & 23:30:32.01  & -10:03:22.14 & n368       & Ia   & Ia-norm    &     83.1 &    100.0 &      0.0 &      0.0 &    0.342 &    0.344 &    0.006 \\ 
---   {\tablenotemark{c}} & 01:13:13.26  & -00:23:25.86 & n400       & ---  & ---        &      --- &      --- &      --- &      --- &    0.424 &      --- &      --- \\ 
---                       & 02:31:31.43  & -08:55:11.52 & n404       & Ia   & Ia-norm    &    100.0 &    100.0 &      0.0 &      0.0 &      --- &    0.216 &    0.008 \\ 
---   {\tablenotemark{c}} & 02:31:19.60  & -08:45:09.76 & n406       & ---  & ---        &      --- &      --- &      --- &      --- &      --- &      --- &      --- \\ 
---                       & 23:29:56.19  & -08:34:24.34 & p425       & Ia   & Ia-norm    &    100.0 &    100.0 &      0.0 &      0.0 &    0.458 &    0.453 &    0.006 \\ 
---   {\tablenotemark{c}} & 01:12:40.25  & +00:14:56.61 & p434       & ---  & ---        &     --- &     61.7 &     33.3 &      4.9 &    0.339 &      --- &      --- \\ 
---                       & 02:08:32.45  & -03:33:34.20 & p454       & Ia   & Ia-norm    &    100.0 &    100.0 &      0.0 &      0.0 &      --- &    0.695 &    0.010 \\ 
---                       & 02:11:00.02  & -04:09:37.59 & p455       & Ia   & Ia-norm    &     88.9 &    100.0 &      0.0 &      0.0 &    0.298 &    0.284 &    0.006 \\ 
---   {\tablenotemark{c}} & 02:08:09.34  & -03:48:05.05 & p520       & ---  & ---        &      --- &      --- &      --- &      --- &      --- &      --- &      --- \\ 
---                       & 02:30:10.16  & -08:52:50.84 & p524       & Ia   & Ia-norm    &    100.0 &    100.0 &      0.0 &      0.0 &      --- &    0.508{\tablenotemark{b}} &    0.009 \\ 
---   {\tablenotemark{c}} & 02:08:10.47  & -03:32:17.70 & p527       & ---  & ---        &     --- &      --- &      --- &      --- &    0.435 &      --- &      --- \\ 
---                       & 02:07:04.66  & -03:28:04.37 & p528       & Ia   & Ia-norm    &     88.2 &    100.0 &      0.0 &      0.0 &    0.781 &    0.777 &    0.005 \\ 
---                       & 02:04:56.09  & -03:49:03.67 & p534       & Ia   & Ia-norm    &     79.1 &    100.0 &      0.0 &      0.0 &    0.619 &    0.615 &    0.008 \\ 
\enddata
\tablenotetext{a}{A secure type was determined, but not a secure
  subtype: there was a majority of correlations with one subtype, but
  the best-match template was of a different subtype.}
\tablenotetext{b}{Only one template which exceeds the cutoff for
  ``good'' correlations: the reported redshift is that of the
  best-match template (as opposed to the median redshift) and the
  associated error is the formal redshift error for that template
  (see \citealt{blondin-snid}).}
\tablenotetext{c}{No ``good'' correlations for this object. No type or
  redshift information is reported.}
\tablenotetext{d}{While there were ``good'' correlations for this
  object, a secure type could not be determined, and we report no
  redshift for this object.}
\tablecomments{
{\bf Column Headings:}  
(1) Official IAU supernova designation; note that not all objects
listed here have official International Astronomical Union names; 
(2) ESSENCE internal identification; 
(3) Supernova type as determined using SNID (see text for details); 
(4) Supernova subtype as determined using SNID (see text for details); 
(5) Absolute fraction of supernova templates corresponding to the
supernova subtype listed in column (4); 
(6) Absolute fraction of supernova templates corresponding to type Ia
supernovae; 
(7) Absolute fraction of supernova templates corresponding to type Ib
or Ic supernovae; 
(8) Absolute fraction of supernova templates corresponding to type II
supernovae;  
(9) Redshift measured from narrow emission or absorption lines from
the host galaxy; 
(10) Redshift as determined using SNID (see text for details); 
(11) Redshift error on the SNID redshift (see text for details).
}
\label{snid_table}
\end{deluxetable}

\pagebreak
\section {Photometry of ESSENCE supernovae}

\subsection{Importance of Photometric Calibration}
\label{sec:calib}

Our ability to determine cosmological parameters from the observations
of supernovae depends on measuring the fluxes of these objects
accurately.  Errors in photometric calibration translate into errors
in the cosmology in two basic ways.  First, we must understand the
calibration of our supernovae fluxes to those of the low-redshift
sample \citep{hamuy93,riess99,jha-lowz06}.  Light curve fitting and
luminosity estimation methods have been trained using these objects
and they also serve the ``anchor'' for the Hubble diagram in our
cosmological measurements of the evolution of the scale factor.
Second, accurate passband-to-passband calibration is important for
estimating the colors of our supernovae, to provide constraints on
extinction due to host galaxy dust.  See the discussion in
\citet{wood-vasey07} for a discussion of how these calibration
issues impact our cosmological measurements.

Photometric systems are defined by the broadband fluxes of a single
standard star (conventionally Vega, though more recently the Sloan
Digital Sky Survey and others have used the F0 subdwarf
$BD+17^{\circ}4708$), as well as a network of standard stars whose
fluxes have been calibrated relative to the primary standard
\citep{landolt83, landolt92}, and the wavelength-dependent
sensitivities of that system.  Observers usually account for the
difference between the particular system they are using and the
standard system by correcting their observations through terms
proportional to the broad-band colors.  These linear corrections can
be quite accurate when derived from observations of standard stars and
then applied to correct the photometry of other stars observed, since
stellar spectra are generally relatively smooth.  However, supernovae
have complex spectra with broad and deep features, and they evolve in
time, so the corrections derived from observations of stars are not
appropriate for calibrating supernova fluxes into a standard system.

To avoid additional error from converting the observed supernova
fluxes to a "standard" system, we report our photometry in the natural
system of the CTIO 4m MOSAIC camera:

\begin{equation}
m=-2.5\log\,\mathcal{F}(ADU)+zeropoint,\end{equation}

where the zeropoints are defined relative to the star Vega.  It is
important to note that in the process of defining a Vega-based
standard star system, the ``true'' magnitudes of Vega have actually
drifted and are slightly non-zero
\citep{bessell-vega98,bohlin-vega04,bohlin-vega06}.  While these
offsets amount to changes in the flux scale of only a few percent,
they become significant for cosmological measurements at the level of
precision we desire and must be accounted for (see
\citet{wood-vasey07} for our treatment of these in the cosmological
analysis).

In the following sections, we describe the calibration of the ESSENCE
photometry in the CTIO 4m natural system.

\subsection {Calibration of ESSENCE field stars}

\begin{figure}
  \plotone{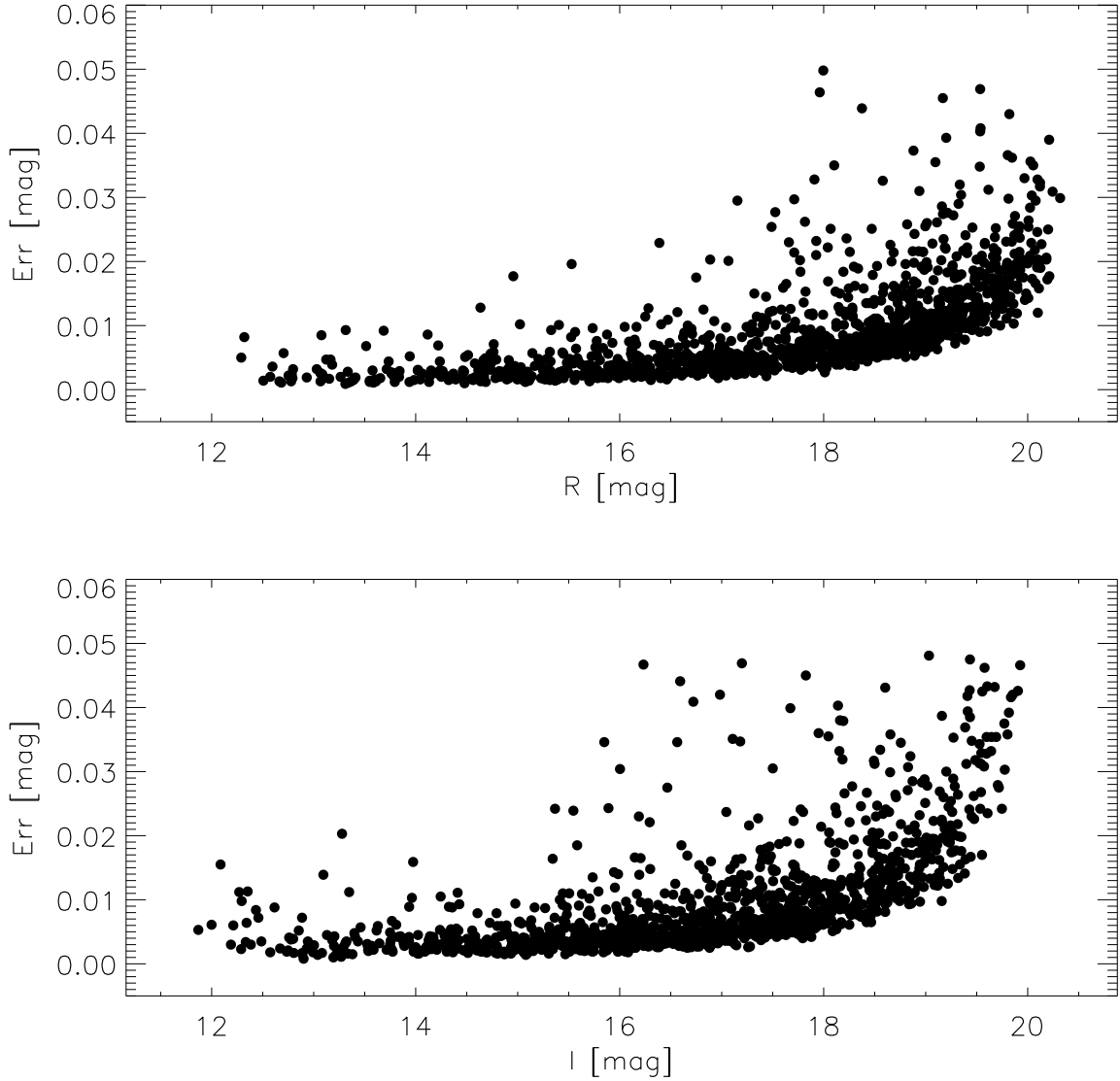}
  \caption{Error in 0.9m photometry of ESSENCE field stars as a function 
    of R magnitudes (top panel) and I magnitude (bottom panel).
    Individual stars are typically measured to a precision 
    of 2\% or better.
    \label{09m_photerror}}
\end{figure}

To establish a Vega-based natural system in our ESSENCE fields, we tie
the stars in these fields to the secondary standards of
\citet{landolt83,landolt92}.  Unfortunately, the overhead in acquiring
a sufficient number of observations of these stars with the MOSAIC
imager is quite high ($\sim$100 seconds readout time, with additional
time spent changing filters and pointing the telescope) relative to
the very short exposures needed to observe these bright objects on a
4m class telescope.  Therefore we have elected to calibrate stars in
our fields with an auxiliary program using the CTIO 0.9m telescope.
Concurrent with the ESSENCE program, we have used 16 photometric
nights on the 0.9m to observe both Landolt standards and ESSENCE field
stars, resulting in 32 calibration patches within the ESSENCE survey.
Each patch contains 40-60 stars observed on a minimum of 3 photometric
nights.  The quality of the photometric calibrations resulting from
the 0.9m program is quite good, with individual stars calibrated to
$\sim$1\% (Figure \ref{09m_photerror}).

\subsection {CTIO 4m photometric zeropoints}

While the 0.9m photometry allows for the transfer of photometric
calibrations in the Vega system to our 4m data, it is not sufficient
to calibrate all of our ESSENCE data, due to the small ($13'$) field
of view relative to the MOSAIC imager.  Each 0.9m patch allows us to
calibrate data from only one of the 8 CCDs in the CTIO MOSAIC.
Therefore, by using our own data taken on photometric nights and
carefully propagating photometric zeropoints from the overlapping data
to rest of the MOSAIC, we generate catalogs which cover our fields
completely.  These catalogs effectively define the ESSENCE photometric
system and are used to calibrate data taken on all other nights.

First, we must transform the 0.9m magnitudes from the system defined
by the Landolt standard stars to the CTIO MOSAIC natural photometric
system, via equations of the form:

\begin{equation}
R_{CTIO}=R_{Landolt}+k_{RI}^{R}(R_{Landolt}-I_{Landolt})
\end{equation}

and

\begin{equation}
I_{CTIO}=I_{Landolt}+k_{RI}^{I}(R_{Landolt}-I_{Landolt}).
\end{equation}

Thus we choose to adopt the same zeropoint for $R_{CTIO},I_{CTIO}$ for
stars of zero color in the Landolt system. The color terms
$k_{RI}^{R},k_{RI}^{I}$ may then be measured by comparing Landolt
standard magnitudes with MOSAIC instrumental magnitudes.

We obtain the values $k_{RI}^{R}=-0.030$ and $k_{RI}^{I}=0.030$ by
combining our own work with the information reported on the CTIO 4m
web page
\footnote{\textit{http://www.ctio.noao.edu/mosaic/ZeroPoints.html}}
and with synthetic photometry using 4m MOSAIC passbands and the
Stritzinger et al. 2005 spectrophotometric standards. These were also
cross-checked by combining aperture corrected DoPHOT magnitudes and
0.9m catalogs.

This 0.9m photometry, now transformed to the CTIO MOSAIC natural
photometric system, was used to compute the zero points of the two
subfields covered by the 0.9m field of view for each of the ESSENCE
fields.  To generate catalogs for the other subfields, we must
propagate the photometric zeropoint from these subfields across the rest
of the MOSAIC. This requires that the instrumental sensitivities
are normalized to a common level, such that one data unit corresponds
to the same amount of incident flux for every subfield, and that we
measure the \emph{same} fraction of the flux for the stars in all the
images, which can be achieved by correcting the PSF magnitudes to an
aperture which encloses the total flux.  If these two conditions are
met, then the zeropoint derived for one subfield is valid for the
entire MOSAIC.

To ensure that the sensitivities are normalized from subfield to
subfield, we use the ratios of the sky levels between subfields, for
all the images for a given night to establish these relative flux
scalings. Because of the enormous numbers of pixels used to measure
this ratio, the results are incredibly precise and the normalization
factors can be measured to $\sim0.3\%$.  This method actually
normalizes the CCD sensitivities for the spectral energy distribution
of the sky, which obviously differs from those of astronomical objects
we seek to measure.  However, tests using a range of passband
sensitivity curves and a variety of input spectra show that the
resulting photometric errors are much less than $1\%$.  Though the
method is unaffected by uniform variations in sky brightness across
the entire MOSAIC, care must be taken to avoid applying this method in
the presence of moonlight which could result in a systematic gradient
in sky brightness across the array.

We then turn to the photometry of stars in the ESSENCE fields, which
have been measured using DoPHOT PSF photometry.  To correct these
magnitudes so that they measure the total flux for the objects in the
images, we use the standard method from aperture photometry of
constructing a {}``growth curve'' for each image from the incremental
flux in concentric annuli about the objects. We choose a small
aperture, for which we robustly determine the offset between the PSF
magnitudes and aperture magnitudes for the brightest stars in the
image.  We then construct a growth curve out to an aperture at large
enough radius that the flux measured at those annuli is consistent
with zero.  Such aperture corrections are calculated for each
subfield-image in a field and are then used to bring all of the PSF
photometry onto the same flux scale.  Note that while the PSF does
vary across the field of view of the MOSAIC, the small number of
isolated stars in a typical ESSENCE field makes robust determination
of spatially varying aperture corrections difficult, so instead a
single correction is calculated for each subfield-image.

\begin{figure}
  \plottwo{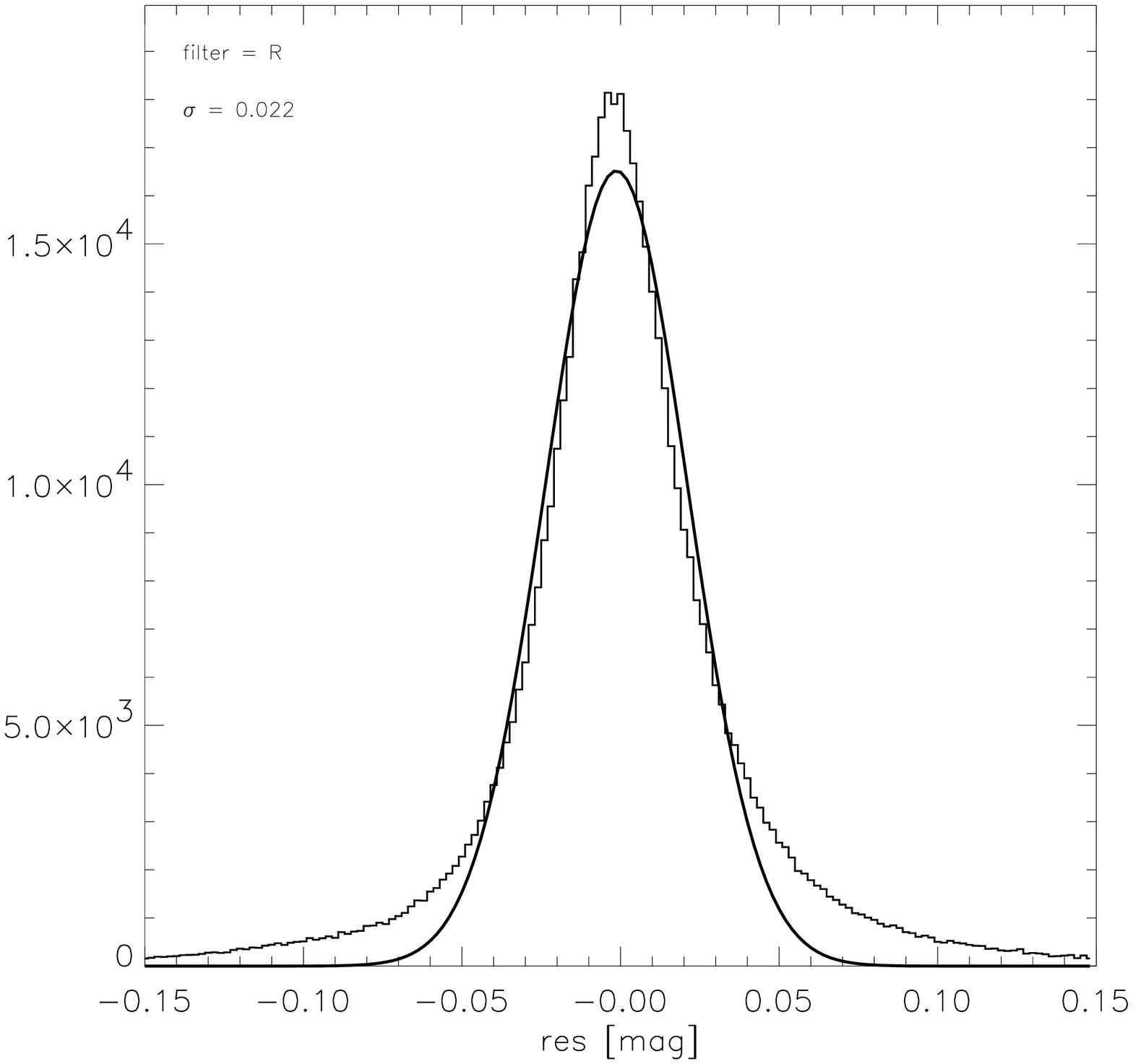}{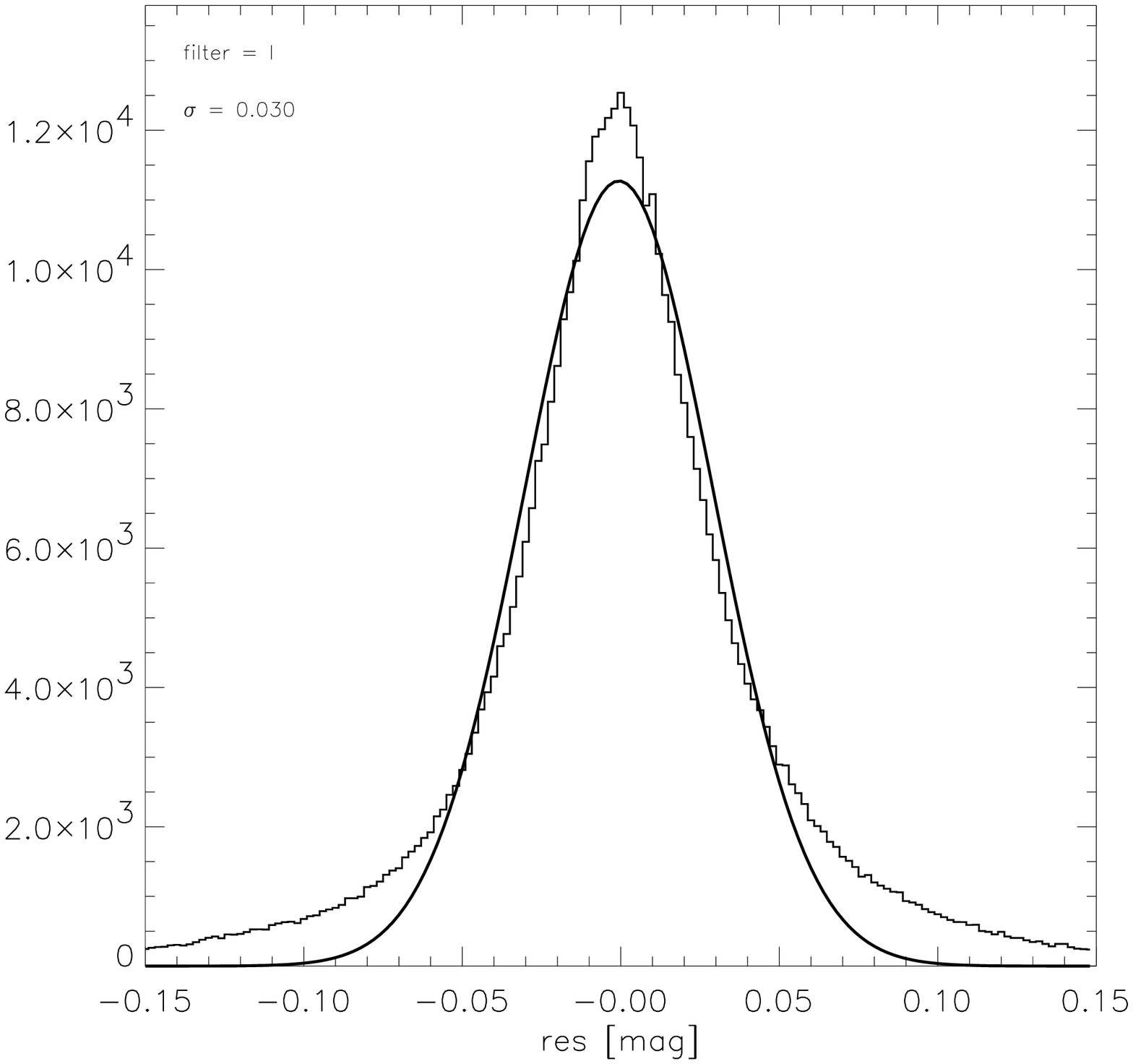}
  \caption{Distribution of the night-to-night photometric residuals in 
    magnitudes for CTIO 4m $R$ (left plot) and $I$ band (right plot) 
    bands for ESSENCE field stars.  The solid lines
    are Gaussians fit to the data.  The small widths of the histograms 
    ($\sim$2-3\%) demonstrates the temporal stability of our photometry.
    \label{residual_hist}}
\end{figure}

With the photometry of the stars in all subfields now on the same flux
scaling, we are able to propagate photometric zero points across the
whole MOSAIC.  In this manner, we calibrate magnitudes for all the
stars present in our fields for several epochs and then compute
$\sigma$-clipped averages over all of the measurements.  Figure
\ref{residual_hist} demonstrates there is a small dispersion in the
residuals about the mean for all the stars in our catalogs.  This
shows that the zeropoint propagation procedure is robust from night to
night.

\begin{figure}
  \plottwo{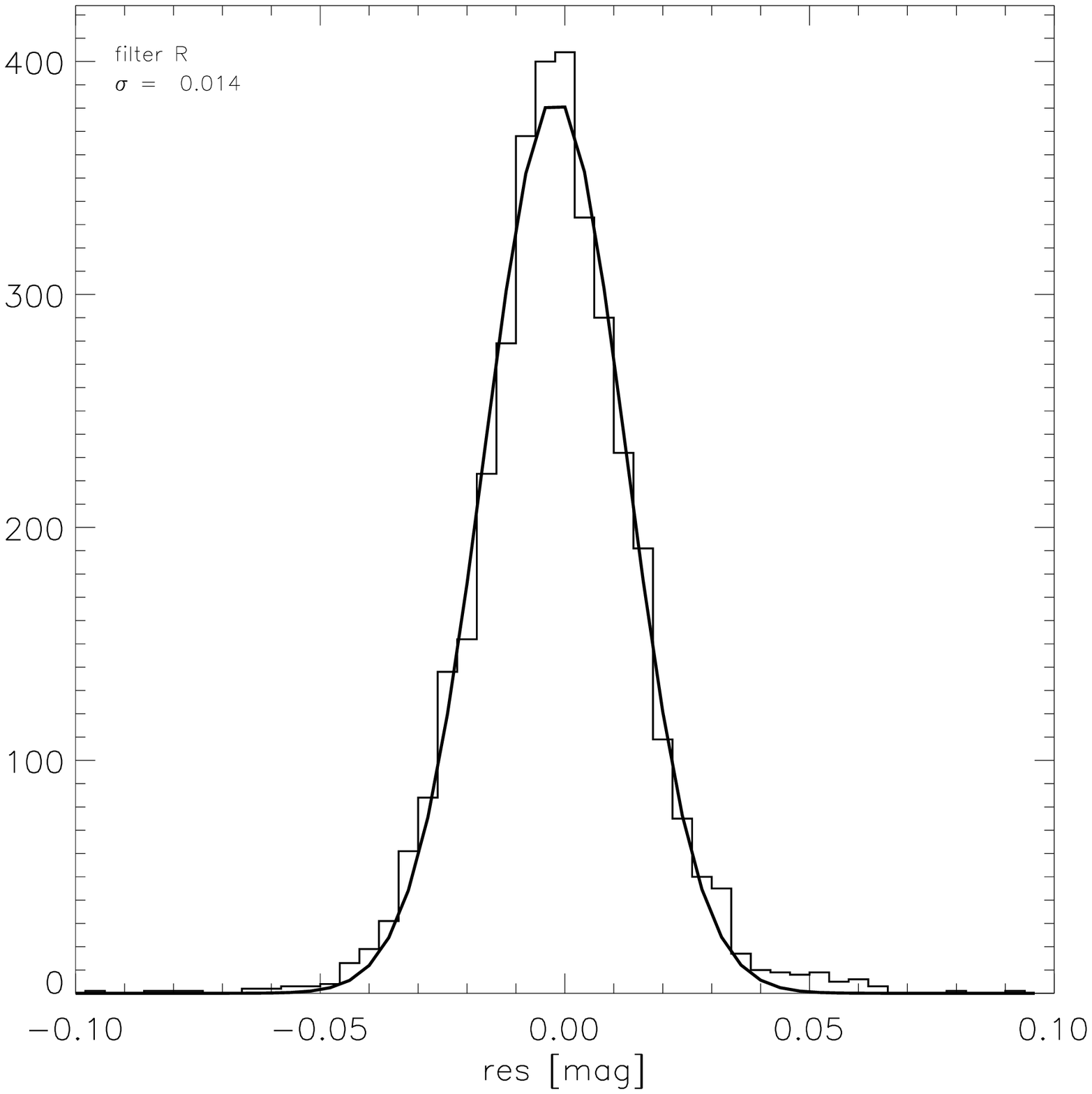}{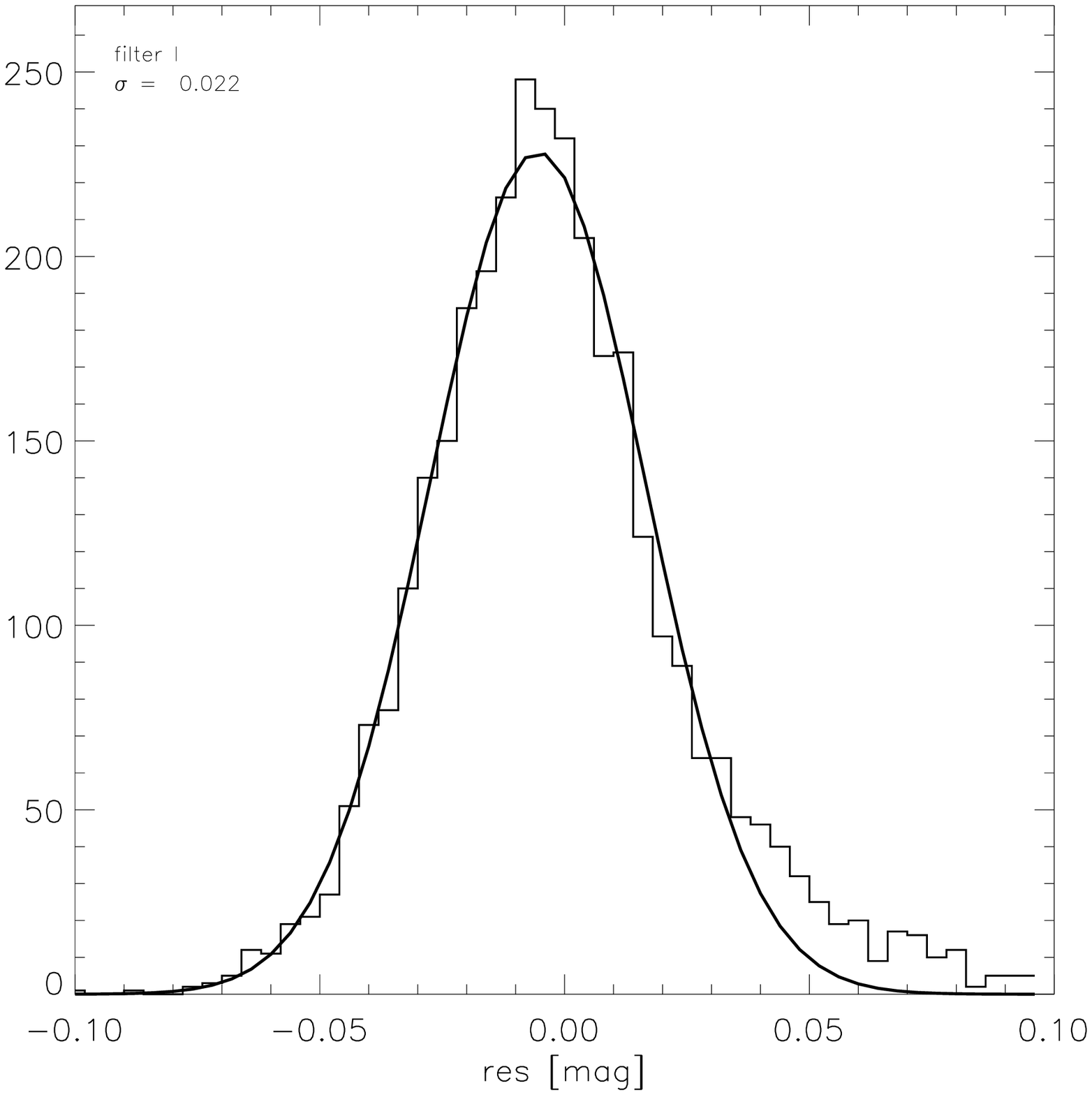}
  \caption{Distribution of photometric zeropoint residuals, in magnitudes,
    for the $R$ (left plot) and $I$ band (right plot) bands.  The small 
    scatter of $\sim$1-2\% demonstrates that our zeropoints are homogenous 
    across the ESSENCE fields. The
    solid lines are Gaussians fit to the data.
  \label{zpt_scatter}}
\end{figure}

To check the field to field consistency of these catalogs, we consider
the ESSENCE data taken under photometric conditions.  For a given
night, we correct the zero points for each by applying aperture and
airmass correction. We then take the average value of those corrected
zero points as the true zeropoint for the entire night.  We then also
calibrate each field individually, using our photometric catalogs.  In
Figure \ref{zpt_scatter}, we show the distribution of the differences 
between the zeropoints calculated using the ESSENCE catalogs and the 
average nightly zeropoint.  The small scatter of $0.02$ magnitudes in each
passband assures us that the zeropoints are consistent from field to field
with a precision of better than 2\%.

\pagebreak

\subsection {Supernova flux measurement}

With accurately determined fluxes of the stars in our fields in the
natural system, we then seek to measure the supernova fluxes as
accurately as possible.  This requires that we remove the background
light due to the host galaxy, via image subtraction using the same
software as in the search pipline (section \ref{imagesub}).  It is
crucial that the subtraction procedure maintains the flux scaling from
the original image, which has been calibrated to stars, through to the
subtracted image, where we measure the supernova flux.  We have
performed extensive tests by adding synthetic stars to our data to
verify that the registration and subtraction processes do not bias the
supernova photometry.

\begin{figure}
  \plotone{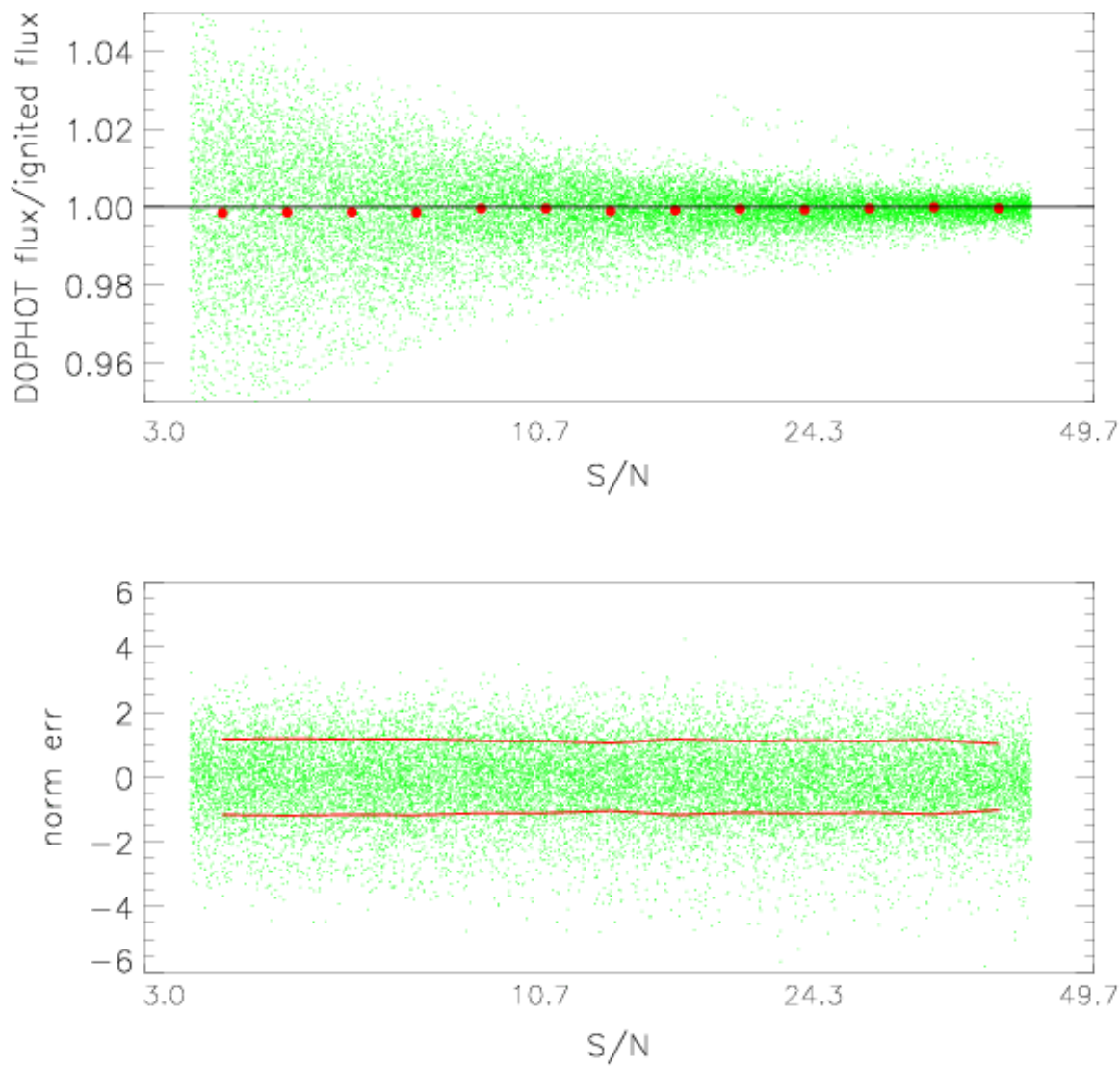}
  \caption{Fake stars were added to images before remapping, which
    rebins pixels.  The top panel shows the ratio of 
    the flux measured to the input flux in the rebinned image, as a 
    function of the
    signal-to-noise ratio of the fake star.  Green points are individual
    stars, red points are averages.
    Rebinning does not significantly
    bias our photometry, even at low SNR.  The bottom panel shows the ratio
    of the flux residuals (input flux - measured flux) divided by the
    estimated error using our noise maps, as a function of SNR.  The
    red lines denote one standard deviation.  We find that the
    distribution is slightly broader than expected ($\sigma=1.1$),
    indicating that our nominal error computed using the noise maps
    slightly underestimates the actual error by 10\%.
    \label{after_swarping}}
\end{figure}

\begin{figure}
  \plotone{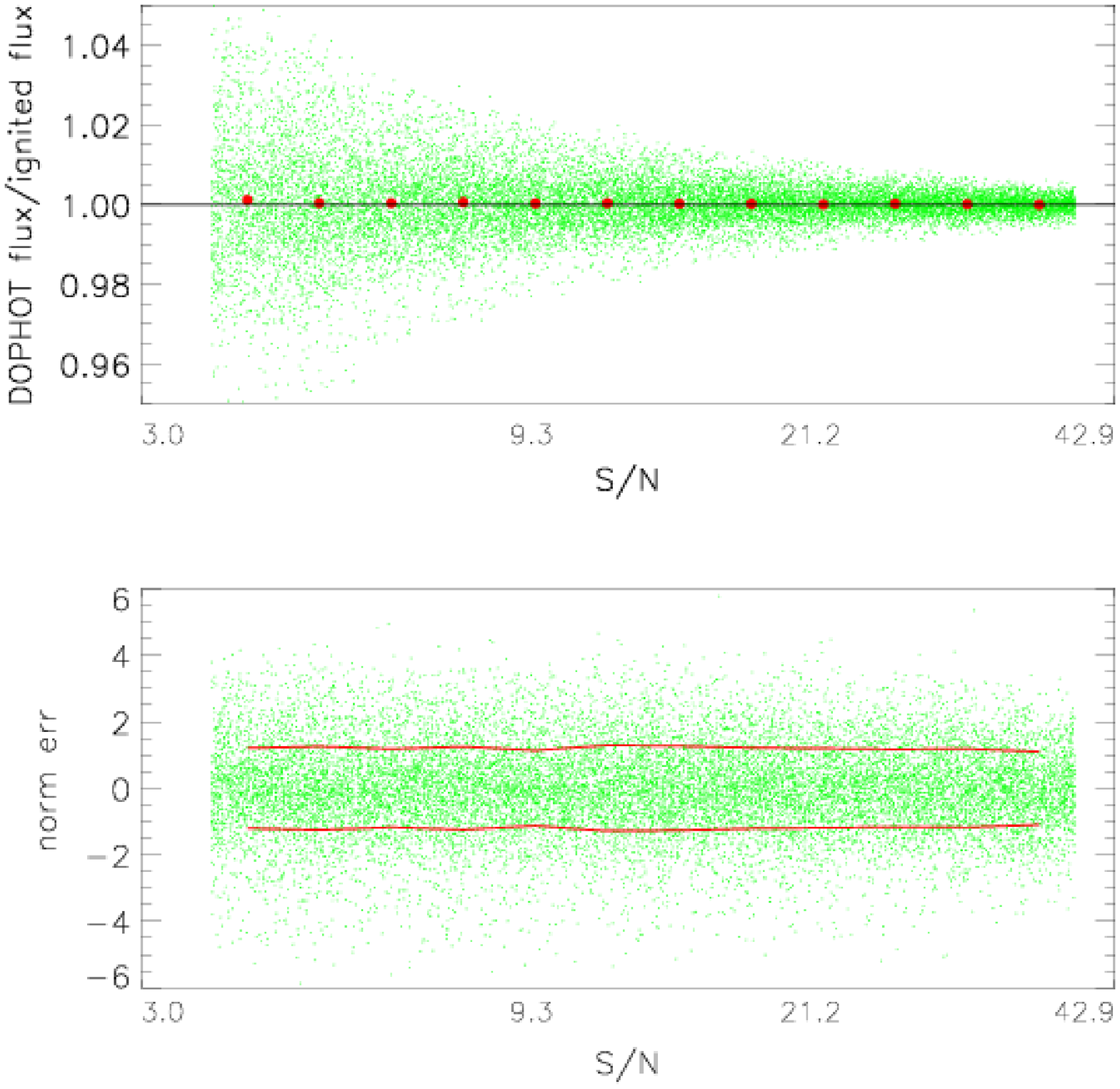}
  \caption{Same as Figure \ref{after_swarping}, except the fake stars
    have been remeasured after template subtraction.  The photometry remains
    linear to much better than 1\%, even at low SNR.  
    The normalized error distribution has 
    $\sigma=1.2$, so we scale the photometric error of measurements on
    subtracted images up by 20\% from the value obtained from the noise maps.
    \label{after_subtraction}}
\end{figure}

To test whether the image registration and subtraction stages affect
our photometry, we added thousands of synthetic stars in a subsample
of our images before these steps. The flux of those stars was then
measured after image registration and after template subtraction
respectively.  The results are shown in Figures \ref{after_swarping}
and \ref{after_subtraction}.  We find that image registration and
subtraction do not significantly bias our photometry, though the
nominal photometric error from our noise maps slightly underestimates
the true photometric error.

To further study the errors in our photometry as estimated using the
noise maps, we measure fluxes using the DoPHOT PSF in a regular grid
across the difference image, where there are no sources of flux.  If
the nominal photometric error were accurate, then we should find that
the distribution of flux/$\sigma_{flux}$ measured with the PSF in
these empty regions should be centered on zero with a $\sigma$ of 1.0.
In practice, we find that this distribution is somewhat broader
($\sigma \sim 1.2$) for a typical difference image.  We interpret this
to mean our errors are slightly underestimated, probably due to
pixel-to-pixel covariances generated in the remapping and convolution
steps that are not accounted for properly in the noise maps.  We scale
up the photometric errors for each difference image by the factor 1.2.

On each difference frame, the PSF used to measure the supernova is
determined using field stars prior to subtraction.  For each
subtraction, we convolve the image with the narrower PSF to match the
broader PSF in the other image, it is this broader PSF which is used
to measure the supernova in the subtraction.  The flux calibration of
that same image, from comparing DoPHOT photometry to the catalogs
described in Section \ref{sec:calib}, is scaled by the normalization
of the subtraction kernel and then then applied to the supernova flux
measured in the difference image.  

\begin{figure}
  \plotone{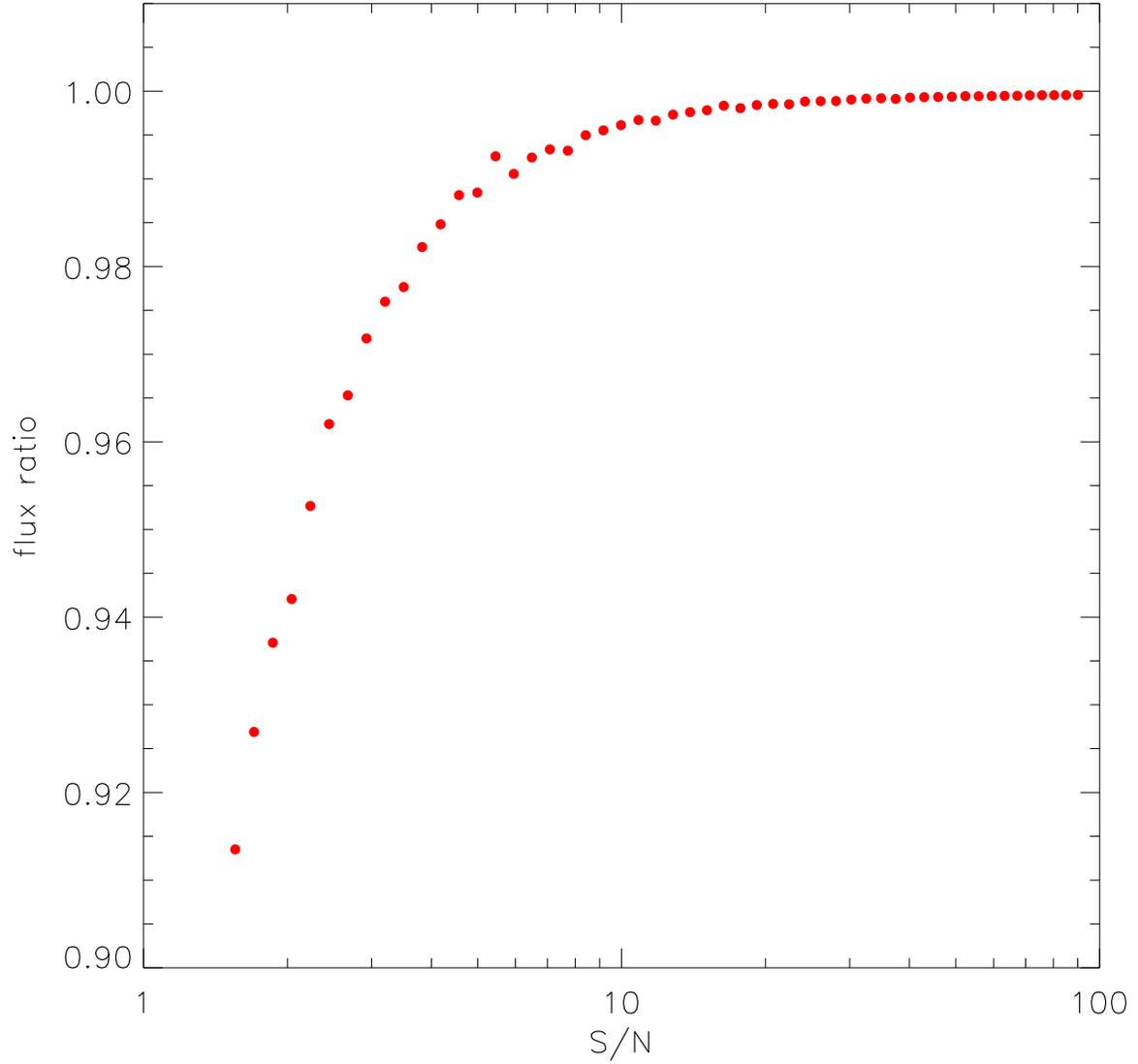}
  \caption{Ratio of recovered to input flux due to systematic
        misalignment of the PSF for the typical centroiding error as
        a function of the cumulative signal-to-noise ratio of the object
        over all photometric measurements.  By combining all measurements
        in both passbands, the positions of even faint SNe are constrained
        at a level corresponding to SNR > 10.
        \label{astrometry_systematic}}
\end{figure}

To measure the supernova flux accurately, we fix the PSF to the best
measured location of the supernova, rather than allow the position to
be a free parameter in the PSF fit.  Because fitting the PSF at a
position displaced from the true source center would result in a
systematic underestimate of the measured flux for the entire light
curve, we estimate the size of this effect for our typical positional
errors.  The location for each supernova is refined from its discovery
position by taking the average of all detections with a
signal-to-noise ratio of 5 or greater in all the available difference
image frames .  These derived positions are accurate to within
$0.02\arcsec$ within our astrometric system.  In
Fig.~\ref{astrometry_systematic} such a systematic is quantified by
artificially shifting sources of known flux that have a FWHM of 1.0
arcsec, the average value for the ESSENCE survey.  Our SN light curves
are usually very well sampled, providing a cumulative signal-to-noise
ratio greater than 10 even for the highest redshift objects.  This
effective signal-to-noise ratio translates to a photometric error less than
1.0\%.

To obtain optimal signal-to-noise in our subtractions, we make use of
all of the images that contain background galaxy light.  We follow the
methodology outlined in \citet{nn2}, which utilizes the flux
differences from all $N(N-1)/2$ possible image pairs to estimate the
supernova flux.

When dealing with the thousands of difference images generated in our
NN2 method, automated and quantitative quality controls were crucial
in extracting good measurements.  A second check was to measure the
flux of known stars in the difference image.  Ideally, there should be
no excess of positive or negative flux in the remaining if the
subtraction process was successful.  After sigma-clipping to reject
variable stars, the average flux/$\sigma_{flux}$ at the positions of
all the stars was measured and if it was inconsistent with the flux
uncertainty expected for the difference image, that difference image
was not used to measure the supernova flux.  Once the
quality-controlled full sets of N*(N-1)/2 data files were generated,
they were run through the nn2 program of \citet{nn2} to generate our
final supernova light curves included in this paper.

\pagebreak
\section {Photometry from the ESSENCE four year sample}

\begin{figure}
  \plottwo{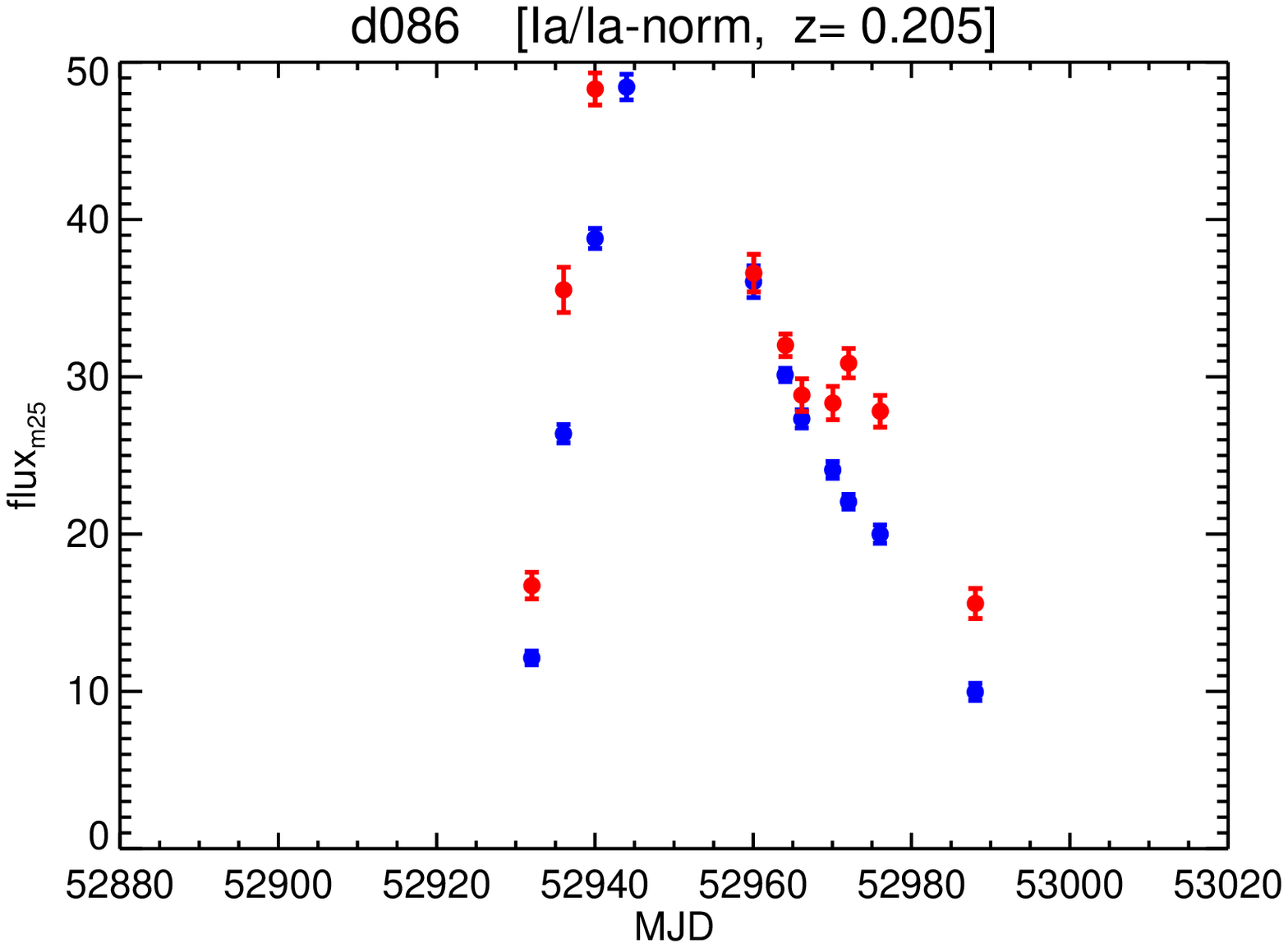}{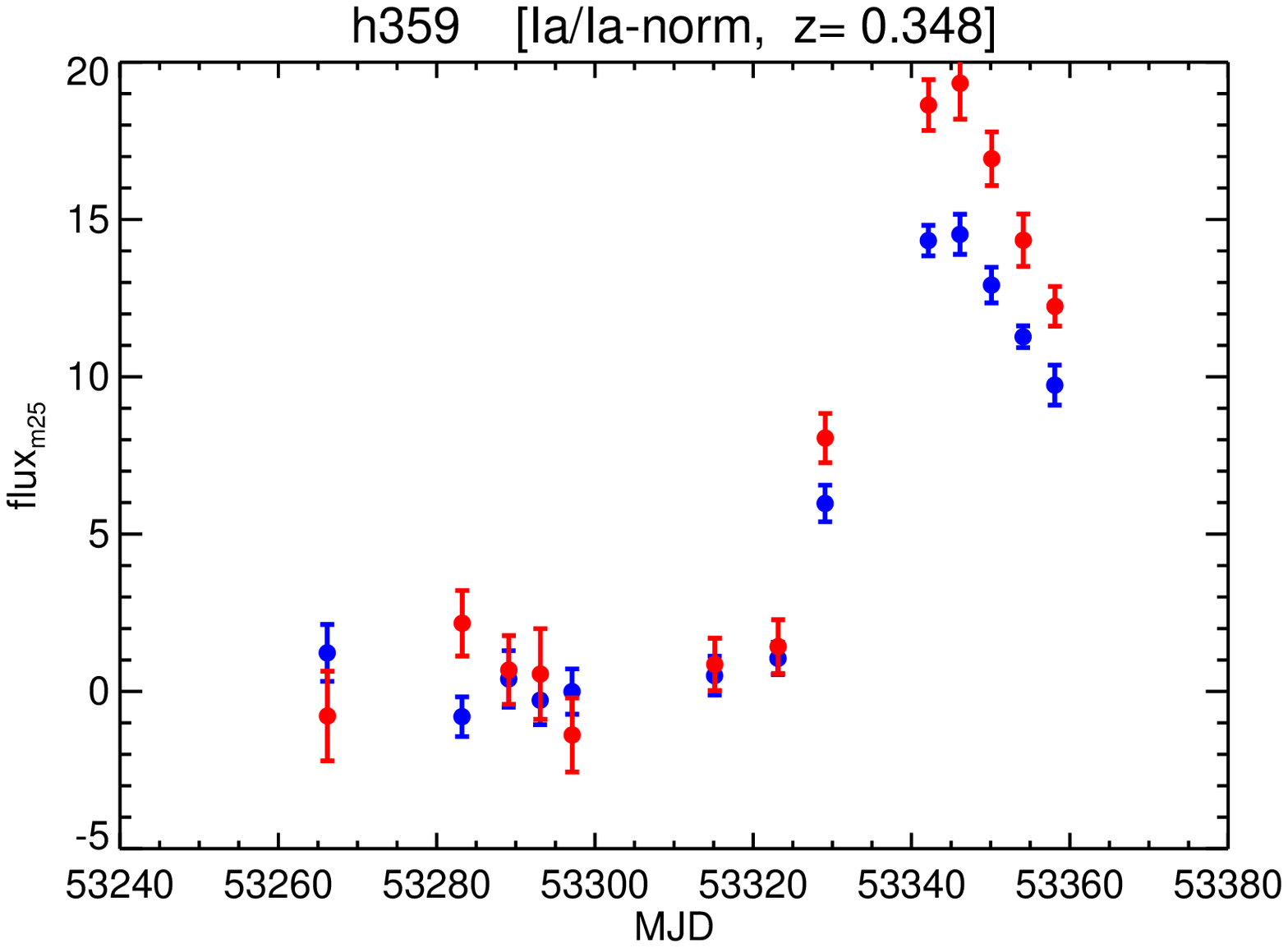}
  \plottwo{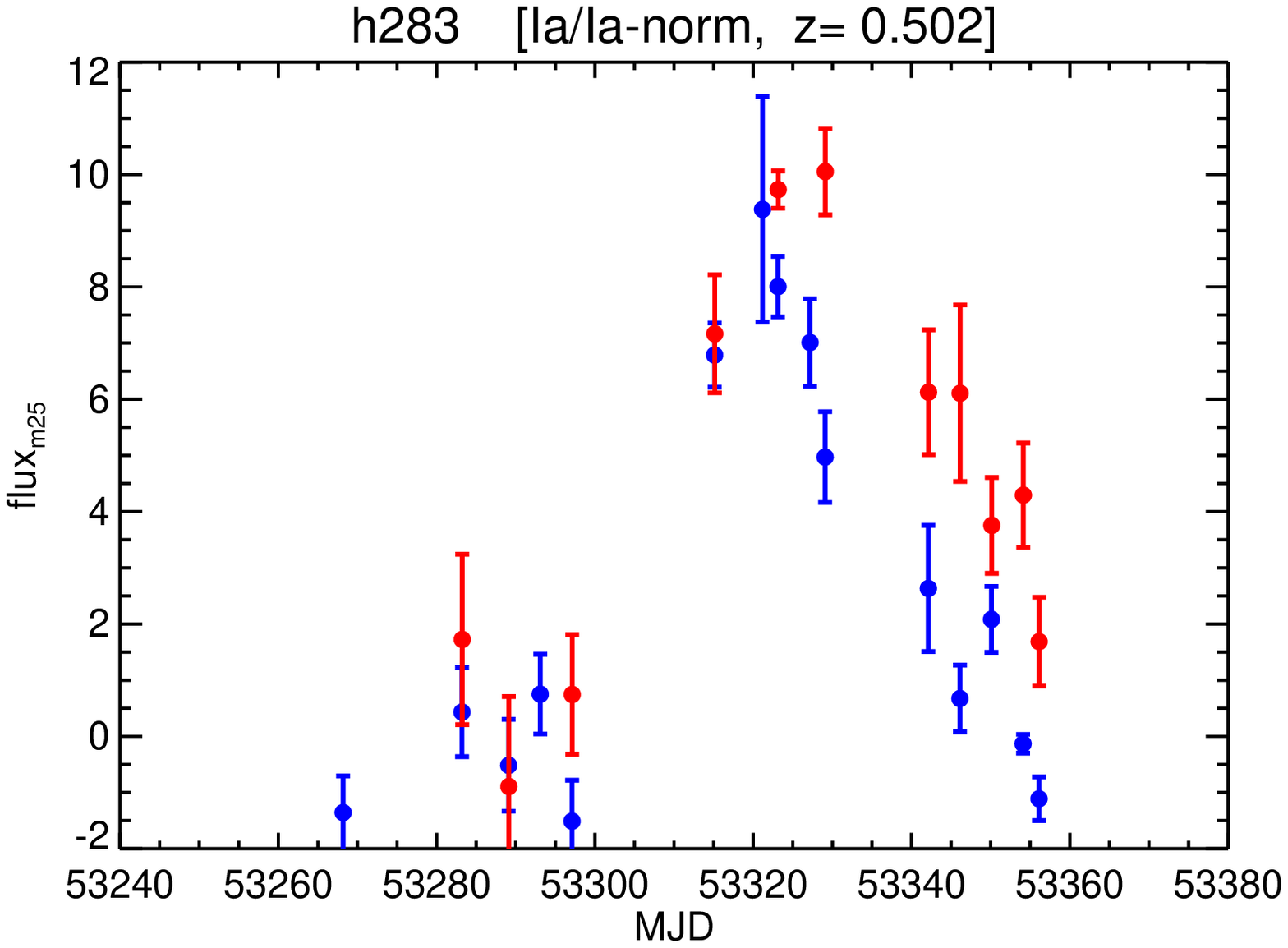}{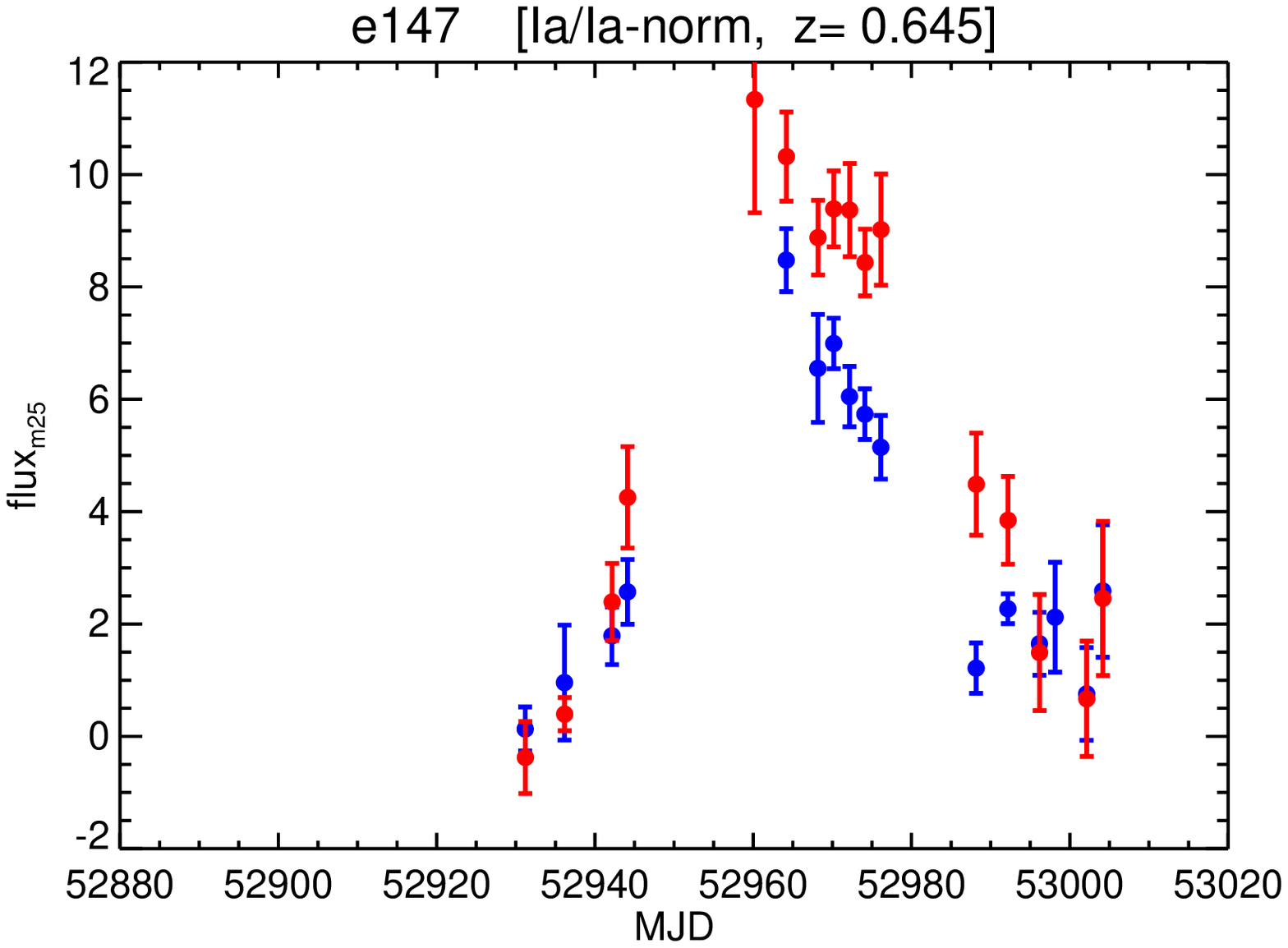}
  \caption{Example ESSENCE light curves, in units of linear flux, 
    scaled such that flux=1 corresponds to magnitude 25 (blue=R, red=I).  
    Only data from the observing season in which the object was 
    discovered are plotted.
    \label{lightcurves}}
\end{figure}

We present here four sample ESSENCE light curve to illustrate the
quality of the ESSENCE photometry (Figure \ref{lightcurves}).  These
objects were chosen to be closest in redshift to an abitrary set of
redshifts, $z=0.20,0.35,0.50,0.65$, which span the range of the
ESSENCE redshift distribution.  For the purposes of plotting, all data
from the season in which the SN was discovered are displayed.
Photometry is presented in linear flux units in the CTIO 4m natural
system, where the formula for conversion to standard magnitudes is

\begin{equation}
m=-2.5\log\,\mathcal{F}+25.\end{equation}

\begin{figure}[th]
\plotone{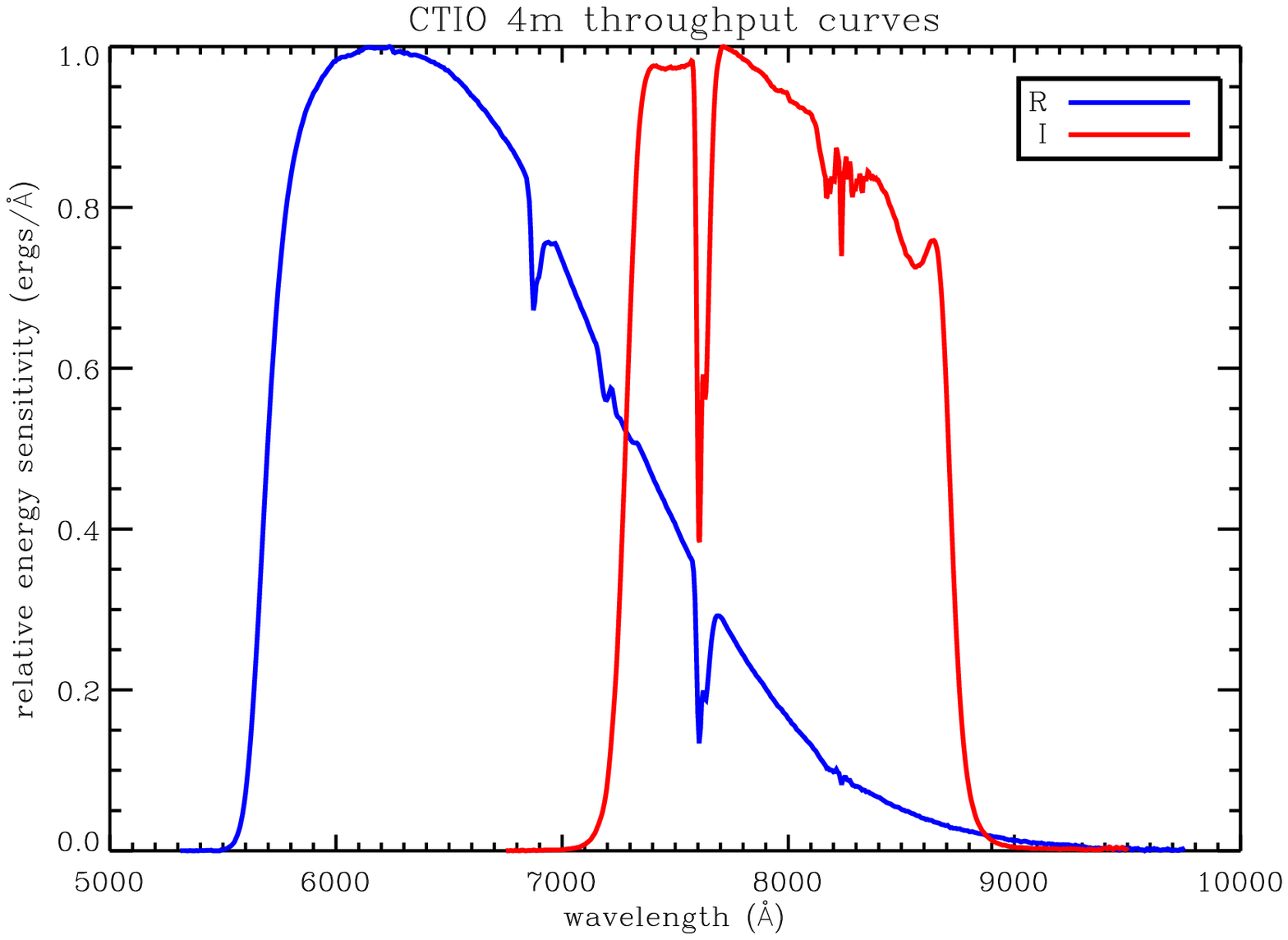}

 \caption{Throughput curves for the CTIO 4m R and I
   bandpasses.  These represent the full system throughput, which
   includes the wavelength-dependence of: the CCD quantum efficiency,
   the optical filters, the aluminum reflectance for the mirrors in
   the 4m telescope and a model for the typical atmosphere
   transmissivity. The curves here are represented in relative energy
   sensitivity in ergs/Angstrom.  Each curve has been normalized to unity 
   at its peak.\label{filters}}
\end{figure}

Since the photometry is reported in the CTIO 4m natural system, the
system throughput curves are an integral part of the data set and are
presented here as well (Figure \ref{filters}).  These system
throughput curves are the product of:

\begin{itemize}
\item the CTIO MOSAIC R and I filters, as measured in the laboratory,
\item standard quantum efficiency curves for the CCDs from the,
  manufacturer (Tek),
\item the wavelength dependence of aluminum, for the two surfaces in
  the 4m telescope and
\item typical atmospheric transmissivity, with losses due to
  scattering and molecular absorption, calculated from taking the
  observations of spectrophotometric standards \citep{hamuy-standards}
  with Bessell's removal of the telluric features
  \citep{bessell-spectrophotometry99} to determine the 
  average atmospheric absorption at CTIO.
\end{itemize}

We are also developing a novel technique for measuring the full
wavelength-dependent response of the telescope/camera system through
the use of a tunable laser and a calibrated photo-diode
\citet{stubbs-calib06}.  Preliminary results from this new method are
consistent with the estimates we derived from the product of each
component as described above.

The full set of ESSENCE light curves and system throughput curves are
available electronically at \url{http://www.ctio.noao.edu/essence/}.

\pagebreak
\section {Conclusion}

We have presented the scientific motivation for the ESSENCE survey,
which aims to constrain the equation of state parameter of dark
energy, $w$, to $10\%$.  Modelling our survey suggests there is a
slight gain in the accuracy of measuring $w$ by covering a greater
volume at lower redshifts by pushing the survey to relatively short
exposure times.  We describe how, using the survey strategy and
software outlined here, we detect likely high-redshift supernovae
using rapid analysis of survey data and how we analyze spectroscopic
data to confidently identify objects as type Ia supernovae and measure
their redshifts.  The photometry for these 102 \snia is presented
here, in the CTIO 4m natural system, as detailed in this document.

Once we have identified the sample of good type Ia supernovae and
carefully measured their light curves, the next step is to estimate
distances to these objects.  A detailed description of the process of
turning supernova photometry and redshifts to cosmological distances
and finally, constraints on cosmological parameters follows in a
companion paper \citep{wood-vasey07}.

ESSENCE has two remaining years of operation.  In addition to
increasing the sample size, we are undertaking a focused effort to
improve the photometric calibration of the CTIO4m and thus reduce the
potential systematic errors from miscalibration.  This program has
been awarded nine nights of engineering time specifically for the goal
of improving the MOSAIC calibrations via concentrated observations of
standard star fields, along with fields observed by ESSENCE and other
on-going CTIO 4m surveys.  With a final sample of $\sim150$ type Ia
SNe and an improvement in photometric precision from the current 2\%
to a final 1\%, we will reach the goal of the project: a measurement
of $w$ to 10\%.

\section{Acknowledgments}

Based in part on observations obtained at the Cerro Tololo
Inter-American Observatory (CTIO), part of the National Optical
Astronomy Observatory (NOAO), which is operated by the Association of
Universities for Research in Astronomy, Inc. (AURA) under cooperative
agreement with the National Science Foundation (NSF); the European
Southern Observatory, Chile (ESO Programmes 170.A-0519 and
176.A-0319); the Gemini Observatory, which is operated by the
Association of Universities for Research in Astronomy, Inc., under a
cooperative agreement with the NSF on behalf of the Gemini
partnership: the NSF (United States), the Particle Physics and
Astronomy Research Council (United Kingdom), the National Research
Council (Canada), CONICYT (Chile), the Australian Research Council
(Australia), CNPq (Brazil) and CONICET (Argentina) (Programs
GN-2002B-Q-14, GS-2003B-Q-11, GN-2003B-Q-14, GS-2004B-Q-4,
GN-2004B-Q-6, GS-2005B-Q-31, GN-2005B-Q-35); the Magellan Telescopes
at Las Campanas Observatory; the MMT Observatory, a joint facility of
the Smithsonian Institution and the University of Arizona; and the F.
L. Whipple Observatory, which is operated by the Smithsonian
Astrophysical Observatory. Some of the data presented herein were
obtained at the W. M. Keck Observatory, which is operated as a
scientific partnership among the California Institute of Technology,
the University of California, and the National Aeronautics and Space
Administration. The Observatory was made possible by the generous
financial support of the W. M. Keck Foundation.

The ESSENCE survey team is very grateful to the scientific and
technical staff at the observatories we have been privileged to use:

{\it Facilities:} 
\facility{Blanco (MOSAIC II)}, 
\facility{CTIO:0.9m (CFCCD)}, 
\facility{Gemini:South (GMOS)}, 
\facility{Gemini:North (GMOS)}, 
\facility{Keck:I (LRIS)},
\facility{Keck:II (DEIMOS, ESI)},
\facility{VLT (FORS1)},
\facility{Magellan:Baade (IMACS)}, 
\facility{Magellan:Clay (LDSS2)}.
  
The survey is supported by the US National Science Foundation through
grants AST-0443378 and AST-057475.  The Dark Cosmology Centre is
funded by the Danish National Research Foundation.  SJ thanks the
Stanford Linear Accelerator Center for support via a Panofsky
Fellowship.  RK thanks the NSF for support through grants AST06-06772
and PHY99-07949 to the Kavli Institute for Theoretical Physics where
he has enjoyed such a splendid sabbatical.  BS, JB and MS thank the
Australian Research Council for support.  This research has made use
of the CfA Supernova Archive, which is funded in part by the National
Science Foundation through grant AST 06-06772. AC acknowledges the
support of CONICYT, Chile, under grants FONDECYT 1051061 and FONDAP
Center for Astrophysics 15010003.

Our project was made possible by the survey program administered by
NOAO, and builds upon the data reduction pipeline developed by the
SuperMacho collaboration.  IRAF is distributed by the National Optical
Astronomy Observatory, which is operated by AURA under cooperative
agreement with the NSF.  The data analysis in this paper has made
extensive use of the Hydra computer cluster run by the Computation
Facility at the Harvard-Smithsonian Center for Astrophysics.  We also
acknowledge the support of Harvard University.  This paper is
dedicated to the memory of our friend and colleague Bob Schommer.

\bibliographystyle{astroads}
\bibliography{apj-jour,ms}

\end{document}